\documentclass[times, twocolumn]{aastex631}

\usepackage{amsmath}
\usepackage{xcolor}
\usepackage{booktabs}

\newcommand{\lya}{Ly$\alpha$}

\newcommand{\ligh}{\textsc{Lightning}}
\newcommand{\chandra}{\textit{Chandra}}
\newcommand{\spitzer}{\textit{Spitzer}}
\newcommand{\herschel}{\textit{Herschel}}
\defcitealias{lehmer2009a}{L09}

\begin{document}

\title{\bf \large Revisiting the Properties of X-ray AGN in the SSA22 Protocluster: \\Normal SMBH and Host-Galaxy Growth for AGN in a $z=3.09$ Overdensity}
\shorttitle{SMBH and Host-Galaxy Growth in the SSA22 Protocluster}

\author[0000-0001-8473-5140]{Erik B. Monson}
\affiliation{Department of Physics, University of Arkansas, 226 Physics Building, 825 West Dickson Street, Fayetteville, AR 72701, USA}
\correspondingauthor{Erik B. Monson}
\email{ebmonson@uark.edu}

\author[0000-0001-5035-4016]{Keith Doore}
\affiliation{Department of Physics, University of Arkansas, 226 Physics Building, 825 West Dickson Street, Fayetteville, AR 72701, USA}

\author[0000-0002-2987-1796]{Rafael T. Eufrasio}
\affiliation{Department of Physics, University of Arkansas, 226 Physics Building, 825 West Dickson Street, Fayetteville, AR 72701, USA}

\author[0000-0003-2192-3296]{Bret D. Lehmer}
\affiliation{Department of Physics, University of Arkansas, 226 Physics Building, 825 West Dickson Street, Fayetteville, AR 72701, USA}

\author[0000-0002-5896-6313]{David M. Alexander}
\affiliation{Centre for Extragalactic Astronomy, Department of Physics, Durham University, South Road, Durham, DH1 3LE, UK}

\author[0000-0001-8618-4223]{Chris M. Harrison}
\affiliation{School of Mathematics, Statistics, and Physics, Newcastle University, Newcastle upon Tyne, NE1 7RU, UK}

\author[0000-0002-7598-5292]{Mariko Kubo}
\affiliation{Astronomical Institute, Tohoku University, Aramaki, Aoba-ku, Sendai, Miyagi 980-8578, Japan}

\author[0000-0002-4933-7208]{Cristian Saez}
\affiliation{Departamento de Astronom\'ia, Universidad de Chile, Casilla 36-D, Santiago, Chile}

\author[0000-0003-1937-0573]{Hideki Umehata}
\affiliation{Institute for Advanced Research, Nagoya University, Furocho, Chikusa, Nagoya 464-8602, Japan}
\affiliation{Department of Physics, Graduate School of Science, Nagoya University, Furocho, Chikusa, Nagoya 464-8602, Japan}

% -----------------------------------------
% Abstract
% -----------------------------------------
\begin{abstract}
We analyze the physical properties of 8 X-ray selected active galactic nuclei (AGN) and one candidate protoquasar system (ADF22A1) in the $z = 3.09$ SSA22 protocluster by fitting their X-ray-to-IR spectral energy distributions (SEDs) using our SED fitting code, \ligh\footnote{\url{https://www.github.com/rafaeleufrasio/lightning}}. We recover star formation histories (SFH) for 7 of these systems which are well-fit by composite stellar population plus AGN models. We find indications that 4/9 of the SSA22 AGN systems we study have host galaxies below the main sequence, with $\rm SFR/SFR_{MS} \leq -0.4$. The remaining SSA22 systems, including ADF22A1, are consistent with obscured supermassive black hole (SMBH) growth in star forming galaxies. We estimate the SMBH accretion rates and masses, and compare the properties and SFH of the 9 protocluster AGN systems with X-ray detected AGN candidates in the Chandra Deep Fields (CDF), finding that the distributions of SMBH growth rates, star formation rates, SMBH masses, and stellar masses for the protocluster AGN are consistent with field AGN. We constrain the ratio between the sample-averaged SSA22 SMBH mass and CDF SMBH mass to $<1.41$. While the AGN are located near the density peaks of the protocluster, we find no statistically significant trends between the AGN or host galaxy properties and their location in the protocluster. We interpret the similarity of the protocluster and field AGN populations together with existing results as suggesting that the protocluster and field AGN co-evolve with their hosts in the same ways, while AGN-triggering events are more likely in the protocluster.
\end{abstract}

% -----------------------------------------
% 1. Introduction
% -----------------------------------------
\section{Introduction}
\label{sec:introduction}

\begin{figure}
\centering
\includegraphics[width=0.5\textwidth]{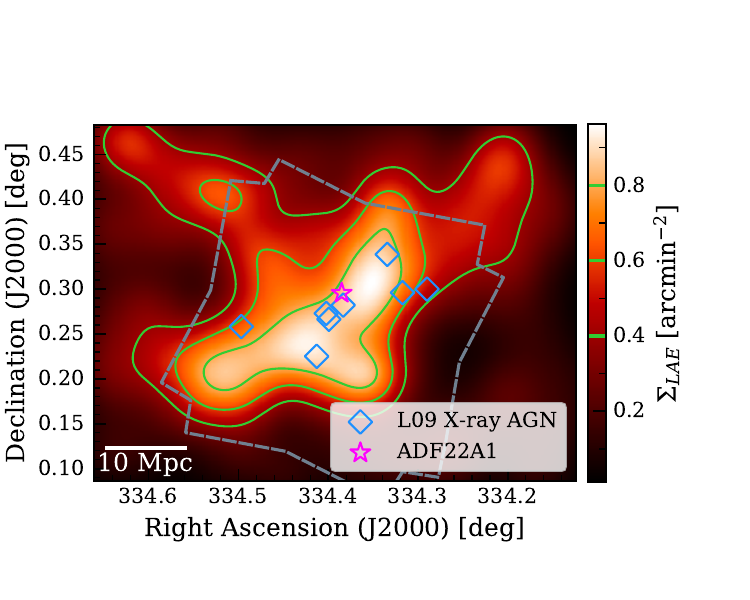}
\caption{\label{fig:SSA22_map}We show the projected density map of $z=3.09$ Lyman-$\alpha$ emitter (LAE) candidates from \citet{hayashino2004} with the locations of the X-ray AGN shown as blue diamonds and the location of ADF22A1 shown as a magenta star. Density contours are also shown at 0.4, 0.6, and 0.8 LAEs $\rm arcmin^{-2}$, and the approximate coverage of the \chandra\ field in SSA22 is shown as a dashed gray line. Three of the X-ray AGN are located within $\sim5$ cMpc of ADF22A1, which is believed to be near the density peak of the protocluster.}
\end{figure}

Galaxies hosting X-ray detected active galactic nuclei (AGN) are observed to be more numerous in the $z \gtrsim 1$ progenitors of galaxy clusters (``protoclusters'') than in local galaxy clusters, where the frequency of X-ray AGN (AGN fraction) is suppressed compared to mean-density (``field'') environments. The enhancement of AGN fraction has now been observed in clusters over $z\approx1$--$1.5$ \citetext{\citealp{martini2013}; see also \citealp{munozrodriguez2023} for a similar result based on semi-empirical models} and numerous protoclusters over $z\approx2$--$4$ \citep[e.g.,][]{digbynorth2010,lehmer2013,vito2020,tozzi2022}, including the $z = 3.1$ SSA22 protocluster, where \citet{lehmer2009a} (hereafter L09) found a $\sim 6$-fold AGN fraction enhancement over the field at the same redshift. The SSA22 protocluster is an intersection of filaments $\sim 60$ co-moving Mpc (cMpc) across at its widest extent (see \autoref{fig:SSA22_map}). It is known to contain an intense overdensity of star forming galaxies \citep[see, e.g.,][]{tamura2009, kubo2013, umehata2015} which are found, along with the AGN, to be coincident with intersections of filamentary cold gas reservoirs in the intergalactic medium (IGM) and Lyman-alpha (\lya) nebulae \citep{umehata2019}. SSA22, then, represents one of our best laboratories for studying how the protocluster environment may drive enhancements of AGN activity.

Observational constraints and the scatter in the AGN fraction--redshift relationship are such that the source of the apparent AGN fraction enhancement in protocluster environments remains uncertain. Simulations of SSA22-like protocluster environments by \citet{yajima2022} suggest that the super-massive black holes (SMBH) in the protocluster grow rapidly in galaxies with $M_{\star} > 10^{10}\ {\rm M_{\odot}}$, potentially reaching masses approaching $10^9\ {\rm M_{\odot}}$ by $z = 3$ before quasar feedback stalls the SMBH growth. The observed AGN fraction enhancement over the field could then be due to larger SMBH masses in the protocluster, as we observe a rapid period of SMBH growth with larger average accretion rates. Alternatively, it could be caused by some modification of the AGN duty cycle in overdense, high-redshift environments, possibly due to more frequent accretion episodes caused by mergers. \citet{monson2021} argued in favor of the former explanation based on finding larger masses among protocluster Lyman-break galaxies (LBGs) compared to field LBGs, and speculated based on non-parametric SFH fitting that the protocluster galaxy population had undergone an earlier period of growth than the field galaxies. However, this sample excluded AGN due to the difficulty of decomposing the stellar population and AGN emission for SED fitting, and hence excluded many of the most massive SSA22 galaxies. In this work, we return to the X-ray detected AGN in SSA22 and fit their SEDs with an improved version of our SED fitting code, \ligh\ \citep{eufrasio2017, doore2021, doore2023}, which now includes models for X-ray to IR AGN emission, allowing us to directly recover SFHs for AGN host galaxies.

The current state-of-the-art SED fitting codes \citetext{e.g., \textsc{CIGALE}, \citealp{boquien2019,yang2020,yang2022}, and \textsc{Prospector-$\alpha$/Prospector}, \citealp{leja2018,johnson2021}} now typically include prescriptions for fitting the continuum emission from an AGN alongside the stellar population of the host galaxy, including, in the case of \textsc{CIGALE} and \textsc{ARXSED} \citep{azadi2023}, X-ray emission. At redshifts where mid-IR observation is difficult, X-ray emission is our primary tool for constraining the AGN luminosity, a necessary step in extracting the SFH of an AGN hosting galaxy. In our updates to \ligh, which we describe in detail in \autoref{sec:sed}, we have added the ability to fit binned X-ray spectra with a physically-motivated X-ray AGN model, which we connect directly to a UV-to-IR AGN model. Our X-ray model allows us to estimate the SMBH masses and accretion rates of the SSA22 AGN, and to connect them to the evolution of the host galaxy.

The paper is organized as follows: \autoref{sec:samples} describes the sample selection and data preparation, \autoref{sec:sed} introduces our models and fitting procedures, \autoref{sec:results} presents our SED fits to the SSA22 X-ray AGN, \autoref{sec:discussion} discusses the implications of our results, and \autoref{sec:summary} summarizes our work.

Where necessary we assume a flat $\Lambda$CDM cosmology with $H_0 = 70\ \rm km\ s^{-1}\ Mpc$, $\Omega_{m,0} = 0.3$, and $\Omega_{\Lambda,0} = 0.7$. In this cosmology, the age of the Universe at $z=3.09$ is 2.04 Gyr. Coordinates are given in the J2000 epoch. Our stellar population synthesis models assume a \citet{kroupa2001} IMF.

% -----------------------------------------
% 2. Samples and Data
% -----------------------------------------
\section{Samples and Data}
\label{sec:samples}

%% -----------------------------------------
%% 2.1 SSA22 Sample
%% -----------------------------------------
\subsection{SSA22}
\label{sec:samples:SSA22}

%%% -----------------------------------------
%%% Table: SSA22 Observations
%%% -----------------------------------------
\begin{deluxetable*}{l r h c c c c c c c r}
\tablecolumns{10}
\tabletypesize{\scriptsize}
%\rotate
\tablecaption{\label{tab:SSA22_obs}Observational properties of the SSA22 sample.}
\tablehead{\colhead{ID} & \colhead{Alt. Name(s)\tablenotemark{a}} & \nocolhead{X-ray Source Number} & \colhead{Redshift\tablenotemark{b}} & \colhead{$\log L_X$\tablenotemark{c}} & \colhead{\textit{Chandra} R.A.\tablenotemark{d}} & \colhead{\textit{Chandra} Dec.\tablenotemark{d}} & \colhead{ALMA 870\micron\ Flux\tablenotemark{e}} & \colhead{ALMA R.A.\tablenotemark{f}} & \colhead{ALMA Dec.\tablenotemark{f}} & \colhead{Association} \\ \colhead{} & \colhead{} & \nocolhead{} & \colhead{} & \colhead{${(\rm erg\ s^{-1})}$} & \colhead{(Deg.)} & \colhead{(Deg.)} & \colhead{(mJy)} & \colhead{(Deg.)} & \colhead{(Deg.)} & \colhead{}}
\startdata
J221736.54+001622.6 & AGN1, ADF22A9  & 140    & 3.084$^{(1)}$           & 44.17 & 334.40225 & 0.2729 & $1.84 \pm 0.21$          & 334.40233 & 0.2729 & \nodata \\
J221739.08+001330.7 & AGN2           & 153    & 3.091$^{(1)}$           & 43.91 & 334.41279 & 0.2252 & $0.91 \pm 0.10$\tablenotemark{*}          & 334.41187 & 0.2260 & LAB2    \\
J221709.60+001800.1 & AGN3           & 20     & 3.106$^{(2)}$           & 44.03 & 334.29025 & 0.2999 & $<1.04$                  & \nodata & \nodata & \nodata \\
J221720.24+002019.3 & AGN4           & 57     & 3.105$^{(2)}$           & 44.61 & 334.33442 & 0.3386 & $<1.10$                  & \nodata & \nodata & \nodata \\
J221735.84+001559.1 & AGN5, ADF22A6  & 139    & 3.094$^{(2)}$           & 44.27 & 334.39933 & 0.2664 & $2.96 \pm 0.29$          & 334.39925 & 0.2664 & LAB14   \\
J221759.23+001529.7 & AGN6           & 268    & 3.096$^{(2)}$           & 44.33 & 334.49675 & 0.2580 & $<1.21\ (1.38 \pm 0.29)$ & 334.49725 & 0.2589 & LAB3    \\
J221716.16+001745.8 & AGN7, ADF22B5  & 43     & 3.098$^{(3)}$           & 43.98 & 334.31721 & 0.2958 & $<1.50\ (2.25 \pm 0.34)$ & 334.31704 & 0.2964 & \nodata \\
J221732.00+001655.6 & AGN8, ADF22A12 & 114    & 3.091$^{(4)}$           & 43.96 & 334.38329 & 0.2821 & $1.58 \pm 0.35$          & 334.38308 & 0.2822 & LAB12   \\
J221732.41+001743.8 & ADF22A1        & 120    & 3.092$^{(5)}$           & 44.33 & 334.38508 & 0.2955 & \nodata                  & \nodata & \nodata & AzTEC1  \\
\enddata
\tablenotetext{a}{AGN are numbered as they appear in \citet{lehmer2009a} and \citet{alexander2016}.}
\tablenotetext{b}{Spectroscopic redshift. Sources are as follows: 1 -- \citet{steidel2003}; 2 -- \citet{matsuda2005}; 3 -- \citet{saez2015}; 4 -- \citet{kubo2015}; 5 -- \citet{umehata2017}.}
\tablenotetext{c}{Rest-frame $2-32$ keV luminosity computed from the \citet{lehmer2009b} $0.5-8.0$ keV flux, assuming the inferred $\Gamma$ from their catalog.}
\tablenotetext{d}{Position of the \textit{Chandra} detection produced by \textsc{ACISExtract}.} 
\tablenotetext{e}{See \citet{alexander2016}, Tables 1 and 2. Fluxes for offset detections are shown in parentheses; upper limits here are given at $4.5\sigma$.}
\tablenotetext{f}{For detected sources we give the position; for non-detected sources with an offset detection, we give the position of the offset detection.}
\tablenotetext{*}{This is the ALMA Band 7 flux from \citet{ao2017}; this AGN is non-detected at the flux limit of the \citet{alexander2016} observations.}
\end{deluxetable*}

% Alexander+(2016) ALMA numbers for AGN2 <0.92\ (1.11 \pm 0.25)

% Of the L09 AGN,
% The LBGs are:
% J221736.54+001622.6 (S03 redshift)
% J221739.08+001330.7 (dual LBG/LAE)

% The LAEs are:
% J221709.60+001800.1 
% J221720.24+002019.3
% J221735.84+001559.1
% J221739.08+001330.7
% J221759.23+001529.7

% J221716.16+001745.8 had no spec-z in the original work and 
% gets its spec-z in Saez 2015

% J221732.00+001655.6 had no spec-z in the original work and 
% got its in Kubo 2015

\begin{figure*}
\centering
\includegraphics[width=\textwidth]{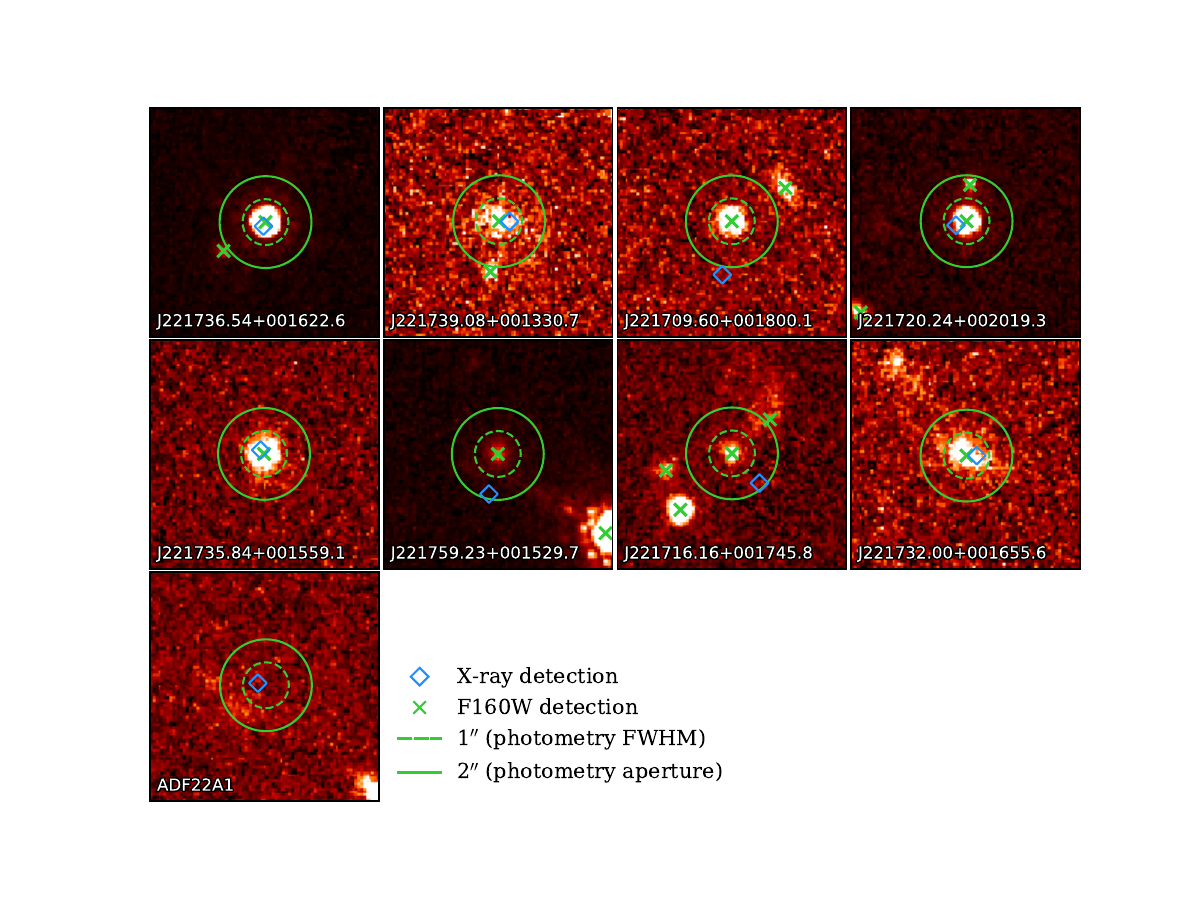}
\caption{\label{fig:SSA22_montage}We show F160W cutouts of the 8 X-ray detected AGN and ADF22A1. The pixel scale of the images as shown is $0.06''$ pix$^{-1}$, and the cutouts are $5''$ on each side. The green dashed and solid circles respectively show the smoothed $1''$ FWHM and $2''$ aperture used to measure the $u^{*}BVR_{c}izJHK_{s}$ photometry from \citet{kubo2013} and F160W photometry from \citet{monson2021}. The \chandra\ detections from \citetalias{lehmer2009a} are shown as cyan diamonds, and $3\sigma$ F160W detections are shown as green crosses. J221720.24+002019.3 has a possible companion within the photometric aperture, which we take as a possible explanation for issues seen with its photometry (see \autoref{sec:samples:SSA22}). Likewise, we see that J2217529.23+001529.7 has a bright foreground star $<4''$ from the main galaxy, which may be blended with the main galaxy in the \spitzer\ IRAC photometry.}
\end{figure*}

Our sample contains the 8 X-ray detected AGN from \citetalias{lehmer2009a} associated with the $z=3.1$ protocluster that were first identified in a 400 ks Chandra survey of the SSA22 field (ObsIDs 8034, 8035, 8036, and 9717). Two of these AGN (J221736.54+001622.6 and J221739.08+001330.7) were first identified as LBGs and spectroscopically confirmed by \citet{steidel2003}; five (J221709.60+001800.1, J221720.24+002019.3, J221735.84+001559.1, J221739.08+001330.7, and J221759.23+001529.7) were detected as $z\approx3.09$ Lyman-$\alpha$ emitters (LAEs) by \citet{hayashino2004} with spectroscopic follow-ups from \citet{matsuda2005}. Of the remaining two systems, J221716.16+001745.8 was spectroscopically confirmed as a protocluster member by \citet{saez2015}, and J221732.00+001655.6 by \citet{kubo2015}. The proto-quasar system ADF22A1, identified as an AzTEC 1.1 mm source in \citet{tamura2010}, is also X-ray detected. We include ADF22A1 in our sample as it possibly represents an early phase of AGN growth in the protocluster, and may (given its projected mass and location in the protocluster) evolve into one of the largest galaxies in the protocluster by $z=0$. The basic observational properties of these 9 sources are given in \autoref{tab:SSA22_obs}. In \autoref{fig:SSA22_montage}, we show F160W cutouts of the X-ray AGN and ADF22A1, along with the positions of the corresponding X-ray detections from \citet{lehmer2009b}. 

For this work the X-ray data were re-reduced using \textsc{CIAO} v4.13 and \textsc{CALDB} v4.9.5\footnote{\url{https://cxc.cfa.harvard.edu/ciao/download/}}, following the procedures described in \citet{lehmer2017}. Source photometry was extracted using the \textsc{ACIS Extract} (\textsc{AE}) pipeline v2020dec17 \citep{broos2010}\footnote{\url{http://personal.psu.edu/psb6/TARA/code/}}, following the procedures outlined in \citet{lehmer2019}. For our SED fits (see \autoref{sec:sed}), we grouped the X-ray photometry into fixed-width bins from $0.5-2.0$, $2.0-4.0$, and $4.0-7.0$ keV. We chose these bins as they were well detected for the majority (6/9) of the sample. In cases where the binomial no-source probability for a bin was $<0.05$, we combined it with the next highest-energy bin, such that the remaining three galaxies have two bins, from $0.5-4.0$ and $4.0-7.0$ keV.

We use UV--NIR photometry from \citet{kubo2013} in the $u^{*}BVR_{c}izJHK_{s}$ bands. These photometry were extracted using a $2''$ diameter circular aperture from images smoothed to a common $1''$ FWHM Gaussian PSF. We also use IRAC channel $3.6\ \micron$, $4.5\ \micron$, $5.6\ \micron$, and $8.0\ \micron$ photometry from \citet{kubo2013}. The IRAC fluxes were not extracted from images smoothed to a common PSF, but were aperture corrected to the same encircled energy as the ground-based photometry. For ADF22A1, we use an additional MIPS $24\ \micron$ measurement calculated by \citet{tamura2010}. We use the \textit{HST} WFC3 F160W fluxes measured in \citet{monson2021}, which were also extracted in a $2''$ diameter circular aperture from images smoothed to a common $1''$ FWHM Gaussian PSF. F160W measurements were available for each of the \citetalias{lehmer2009a} X-ray AGN, but not ADF22A1, which is non-detected at the $S/N$ threshold of the \citet{monson2021} catalog. 

All of our SSA22 sources of interest have been observed at mm/sub-mm wavelengths. Four of the \citetalias{lehmer2009a} AGN and ADF22A1 are detected in ALMA Band 6 ($\sim 1.1$ mm) observations from the ALMA Deep Field in SSA22 (ADF22) survey \citep{umehata2014,umehata2015,umehata2017}. \citet{alexander2016} targeted the 8 X-ray AGN from \citetalias{lehmer2009a} with ALMA Band 7 ($\sim 870\ \micron$), in which 3 of the AGN are detected and the remaining 5 are given 4.5$\sigma$ upper limits (see \autoref{tab:SSA22_obs}). We use these fluxes and upper limits in our SED fits to those sources. Three of the ALMA Band 7-undetected SSA22 AGN (J221739.08+001330.7, J221759.23+001529.7, and J221716.16+001745.8) have associated, offset ALMA detections from \citet{alexander2016}. In subsequent ALMA Band 7 observations by \citet{ao2017}, a source coincident with the J221739.08+001330.7 X-ray source was detected, indicating that the low-S/N offset source may be spurious. We use the \citet{ao2017} flux in our fit to J221739.08+001330.7. For the other two AGN with offset detections, we do not account for the offset source in our fits, instead adopting the upper limit associated with the \chandra\ position. However, in the case of J221716.16+001745.8 we find that the offset detection may be closely associated with the AGN (see \autoref{sec:sed:fitting} and \autoref{app:AGN7_alternative}).

We identified photometric issues with two of the AGN, J221720.24+002019.3 and J2217529.23+001529.7. J221720.24+002019.3 has $J$, $H$, and $K_s$ measurements inconsistent with the rest of the observed SED, producing a very flat spectral slope that we cannot reproduce even with our AGN models. These inconsistencies could be caused by line emission (including possibly broad lines; this source is one of three reported Type 1 AGN in the SSA22 sample), but we also find it likely that these measurements are affected by a nearby projected companion, which is inside the photometric aperture for J221720.24+002019.3. The redshift of the projected companion is unknown. We find also that the ground-based $H$-band measurement is larger than the F160W measurement from \citet{monson2021} by a factor of $\sim 4$. The F160W photometry was measured from an image smoothed to the same FWHM and in the same photometric aperture as the ground-based $H$-band measurement, but the companion may be blended with the AGN in the original, lower-resolution ground-based image, accounting for the difference in flux.

J2217529.23+001529.7 has \spitzer\ IRAC $3.6\ \micron$ and $4.5\ \micron$ measurements that are also inconsistent with the remainder of the observed SED. The IRAC 1 and 2 fluxes are very bright; the IRAC $3.6\ \micron$ flux is the largest in our sample. However, the IRAC colors are inconsistent with AGN emission. The decrement we see between the IRAC $3.6\ \micron$ and $4.5\ \micron$ fluxes is more consistent with stellar emission, and this AGN also has a bright foreground star within $4''$. We suspect that this foreground star may be blended with the AGN in the IRAC data, given the large IRAC PSF. 

Out of an abundance of caution, we treat the measurements listed above for J221720.24+002019.3 and J2217529.23+001529.7 as upper limits in our fits. 

%%% -----------------------------------------
%%% Figure: SSA22 LAE density map
%%% -----------------------------------------

%%% -----------------------------------------
%%% Figure: SSA22 Montage
%%% -----------------------------------------

We correct all of the fluxes for Galactic extinction using the standard \citet{fitzpatrick1999} Milky Way extinction curve and the Galactic dust extinction map from \citet{schlafly2011}, retrieved from the IRSA \textsc{DUST} web application\footnote{\url{https://irsa.ipac.caltech.edu/applications/DUST/}}. We add instrumental calibration uncertainties to the measured uncertainties for each band, along with a flat 10\% model uncertainty in each band. We summarize the bands used for the SSA22 AGN and the fractional calibration uncertainties adopted in \autoref{tab:SSA22bands}.

%%% -----------------------------------------
%%% Table: SSA22 bandpasses
%%% -----------------------------------------
\begin{deluxetable}{l c c}
\tablecolumns{3}
\tablecaption{\label{tab:SSA22bands}Bandpasses and calibration uncertainties used for fits to SSA22 AGN.}
\tablehead{\colhead{Observatory/Instrument} & \colhead{Bandpass} & \colhead{$\sigma_{\rm cal}^{f}$\tablenotemark{a}}}
\startdata
SUBARU/SuprimeCam & $u^*$     & 0.05 \\
\ldots            & $B$       & 0.05 \\
\ldots            & $V$       & 0.05 \\
\ldots            & $R_{c}$   & 0.05 \\
\ldots            & $i'$      & 0.05 \\
\ldots            & $z'$      & 0.05 \\
SUBARU/MOIRCS     & $J$       & 0.05 \\
\ldots            & $H$       & 0.05 \\
\ldots            & $K_{s}$   & 0.05 \\
\textit{HST}/WFC3     & F160W & 0.02 \\
\textit{Spitzer}/IRAC & $3.6\ \micron$ & 0.05 \\
\ldots                & $4.5\ \micron$ & 0.05 \\
\ldots                & $5.6\ \micron$ & 0.05 \\
\ldots                & $8.0\ \micron$ & 0.05 \\
\textit{Spitzer}/MIPS\tablenotemark{*} & $24\ \micron$  & 0.05 \\
ALMA			  & Band 7    & 0.05 \\
\ldots            & Band 6    & 0.05 \\
\enddata
\tablenotetext{a}{Calibration uncertainty as a fraction of flux: $\sigma_{\rm total} = \sqrt{\sigma^2 + (\sigma_{\rm cal}^{f} f)^2}$}
\tablenotetext{*}{ADF22A1 only}
\end{deluxetable}

%% -----------------------------------------
%% 2.2 Deep Field Samples
%% -----------------------------------------
\subsection{Chandra Deep Fields}
\label{sec:samples:CDF}

In order to generate a comparison sample for the SSA22 AGN, we utilize the 2 Ms Chandra Deep Field North (CDF-N) catalog \citep{xue2016} and 7 Ms Chandra Deep Field South (CDF-S) catalog \citep{luo2017}. To select candidate AGN in the $z\sim3$ field, while accounting for the differing X-ray depth between SSA22 and the \chandra\ deep fields, we converted the rest-frame $0.5-7.0$ keV $L_{\rm X}$ in the deep field catalogs to a rest-frame $10-30$ keV luminosity, assuming a photon index $\Gamma = 1.8$. We then selected sources with $L_{10-30\ \rm keV} \geq 2 \times 10^{43}\ \rm erg\ s^{-1}$, the three-count detection limit from the SSA22 X-ray survey of \citet{lehmer2009b}. We cross matched the Cosmic Assembly Near-Infrared Deep Extragalactic Legacy Survey (CANDELS) counterparts given in the CDF catalogs for these candidate AGN with the CANDELS photometric catalogs in the GOODS-N \citep{barro2019} and GOODS-S \citep{guo2013} fields to retrieve optical-NIR photometry. We find 93 candidate AGN with unique photometric matches in GOODS-N, and 107 in GOODS-S.

We also included \herschel\ PACS and SPIRE IR photometry from the CANDELS catalogs \citep{barro2019}, where available. At $z \sim 3$, blending of far-IR detections with \herschel\ PACS and SPIRE is a significant concern. \citet{barro2019} adopt a prior-based extraction method for their FIR fluxes to mitigate this. To be conservative, we additionally searched for \spitzer\ MIPS 24 \micron\ detections within a 1 FWHM diameter aperture around each \herschel\ detection, in each band (the diameters are respectively $7.0''$, $11.2''$, and $18.0''$ for PACS 100 \micron, PACS 160 \micron, and SPIRE 250 \micron). We then calculated the total 24 \micron\ flux within the 1 FWHM aperture and the fraction of the total 24 \micron\ flux attributed to the primary source most closely matched to the \herschel\ detection. If this fraction is less than 0.50, we suspect that the \herschel\ photometry may be influenced by blending, and exclude it from our fits. This process has the side effect of requiring a MIPS 24 $\micron$ detection\footnote{Given our previous selection for highly luminous AGN, the requirement of a 24 \micron\ detection does not appear to introduce significant bias in the X-ray luminosity, AGN obscuration (using hardness ratio as a proxy), or $K-$band luminosity of the sample. The distributions of these observational properties before and after imposing the 24 \micron\ requirement are statistically consistent.}, which leaves 73 AGN candidates in the CDF-N and 76 in the CDF-S. We removed galaxies from the sample for which all three \herschel\ bands are blended, as this gives us no constraint on the dust emission from the galaxy. This final requirement excludes 1 galaxy from the CDF-N. Our sample thus includes 72 candidate AGN in the CDF-N and 76 in the CDF-S.

In the interest of not biasing our sample toward only high-SFR galaxies, we do not require \herschel\ detections for CDF AGN candidates. For AGN with no \herschel\ far-IR detections (i.e., no non-blended detections at observed-frame wavelengths $\geq 100 \micron$), we calculated PACS upper limits from the PEP/GOODS-\herschel\ DR1 \citep{magnelli2013} error maps\footnote{Retrieved from \url{https://www.mpe.mpg.de/ir/Research/PEP/DR1}}, using the optimum circular apertures and aperture corrections from \citet{perezgonzalez2010}. We calculated our own upper limits rather than using published limiting fluxes for the fields due primarily to the variation in PACS depth across the GOODS-S field, which causes the published upper limits to be too restrictive for our purposes for galaxies outside the central, ultra-deep region of the field.

\begin{figure}
\centering
\includegraphics[width=0.5\textwidth]{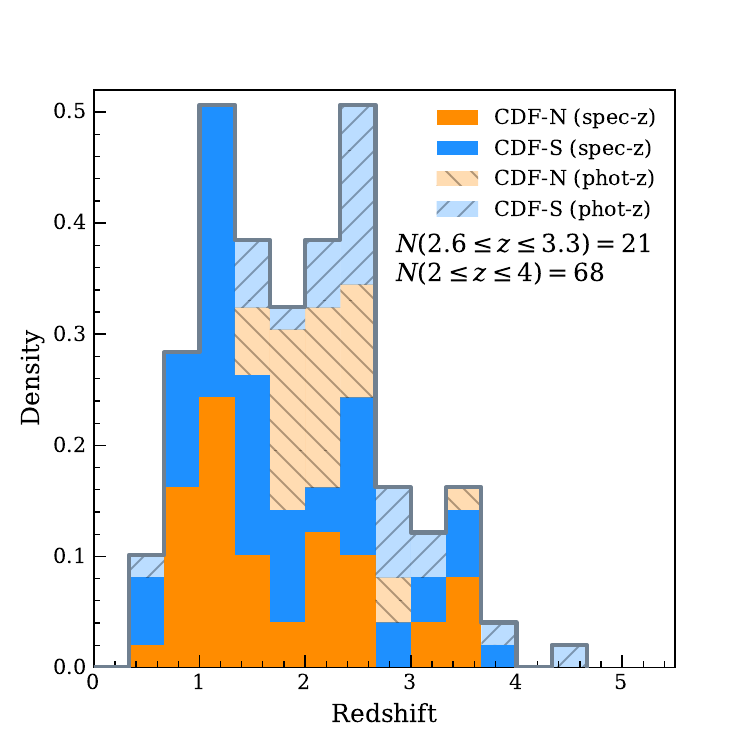}
\caption{\label{fig:GOODS_redshift}We show a stacked histogram of the CDF sample redshifts: solid bars represent spectroscopic redshifts, while desaturated and hatched bars represent photometric redshifts. Including photometric redshifts in our sample improves our coverage at high redshift, crucial to constructing a comparison for the $z=3.09$ protocluster.}
\end{figure}

To fit the SEDs of the CDF AGN, we require redshifts. Of the CDF-N (CDF-S) AGN, 45/72 (52/76) have reliable spectroscopic redshifts (quality flag $\leq 2$ in the CANDELS catalogs). To improve the redshift coverage of our sample we use the photometric redshifts reported in the \citet{guo2013} and \citet{barro2019} catalogs where spectroscopic redshifts are unavailable or reported to be unreliable. The \citet{barro2019} photometric redshifts have three quality tiers, depending on whether they were derived from the broad-band CANDELS photometry (tier 3), CANDELS+SHARDS photometry (tier 2), or CANDELS+SHARDS and grism data (tier 1). Of the 27 CDF-N galaxies in our sample for which we use photometric redshifts, 7 galaxies are at tier 3, 4 at tier 2, and 16 at tier 1. The \citet{guo2013} photometric redshifts are all derived from broad-band photometry, equivalent to the tier 3 redshifts for CDF-N. We estimate the accuracy of these redshifts by computing $| \Delta z | / (1 + z_{\rm spec}) = |z_{\rm phot} - z_{\rm spec}| / (1 + z_{\rm spec})$ for the AGN candidates with spectroscopic redshifts. We find that the outlier fraction $\eta$ of AGN with $| \Delta z |/ (1 + z_{\rm spec}) > 0.15$ is $\eta = 31.1\%$ in the CDF-N sample and $\eta = 26.9\%$ in the CDF-S sample, a factor of 6--8 larger than the outlier fractions reported for the full catalogs in \citet{barro2019} and \citet{guo2013}, respectively, or approximately $28.9\%$ for the full sample. The outlier fraction is relatively stable with redshift, increasing slightly to $32.4\%$ at $z > 2$. We show the distribution of redshifts for each CDF sample in \autoref{fig:GOODS_redshift}. The full sample of CDF-N AGN ranges from $z=0.51$--$3.65$, while the CDF-S AGN range from $z=0.45$--$4.52$. Below, we focus on galaxies with $z \geq 2$ when making comparisons to the protocluster; at these high redshifts the spectroscopic redshift fractions are lower: 17/33 for CDF-N and 17/36 for CDF-S. In what follows we distinguish between galaxies with spectroscopic and photometric redshifts where relevant. We note that despite the large estimated outlier fractions for the spectroscopic redshifts, we do not see systematic differences in the SED-fit derived properties for AGN candidates with spectroscopic and photometric redshifts.

Hereafter we refer to these samples as our CDF-N and CDF-S samples. As with SSA22 (\autoref{sec:samples:SSA22}), we added calibration uncertainties and a 10\% model uncertainty to the measured uncertainties for each band. We summarize the calibration uncertainties for each field and bandpass in \autoref{tab:CDFbands}. 

%%% -----------------------------------------
%%% Table: GOODS bandpasses
%%% -----------------------------------------
\begin{deluxetable*}{l l c c l l c c}
\tablecolumns{8}
\tablecaption{\label{tab:CDFbands}Bandpasses and calibration uncertainties used for fits to CDF AGN candidates.}
\tablehead{\colhead{Sample} & \colhead{Observatory/Instrument} & \colhead{Bandpass} & \colhead{$\sigma_{\rm cal}^{f}$\tablenotemark{a}} & \colhead{Sample} & \colhead{Observatory/Instrument} & \colhead{Bandpass} & \colhead{$\sigma_{\rm cal}^{f}$}}
\startdata
%GOODS-N & KPNO 4m/Mosaic          & $U$            & 0.05 & GOODS-S & Blanco/Mosaic II        & $U$            & 0.05 \\
CDF-N & LBT/LBC                 & $U$            & 0.10 & CDF-S & VLT/VIMOS               & $U$            & 0.10 \\
      & HST/ACS                 & F435W          & 0.02 &       & HST/ACS                 & F435W          & 0.02 \\
      & \ldots                  & F606W          & 0.02 &       & \ldots                  & F606W          & 0.02 \\
      & \ldots                  & F775W          & 0.02 &       & \ldots                  & F775W          & 0.02 \\
      & \ldots                  & F814W          & 0.02 &       & \ldots                  & F814W          & 0.02 \\
      & \ldots                  & F850LP         & 0.02 &       & \ldots                  & F850LP         & 0.02 \\
      & HST/WFC3 IR             & F105W          & 0.02 &       & HST/WFC3 IR             & F098M          & 0.02 \\
      & \ldots                  & F125W          & 0.02 &       & \ldots                  & F105W          & 0.02 \\
      & \ldots                  & F140W          & 0.02 &       & \ldots                  & F125W          & 0.02 \\
      & \ldots                  & F160W          & 0.02 &       & \ldots                  & F160W          & 0.02 \\
      & CFHT/WIRCam             & $K_s$          & 0.05 &       & VLT/HAWK-I              & $K_s$          & 0.05 \\
%      & SUBARU/MOIRCS           & $K_s$          & 0.05 &       & VLT/ISAAC               & $K_s$          & 0.05 \\
      & \textit{Spitzer}/IRAC   & $3.6\ \micron$ & 0.05 &       & \textit{Spitzer}/IRAC   & $3.6\ \micron$ & 0.05 \\
      & \ldots                  & $4.5\ \micron$ & 0.05 &       & \ldots                  & $4.5\ \micron$ & 0.05 \\
      & \ldots                  & $5.6\ \micron$ & 0.05 &       & \ldots                  & $5.6\ \micron$ & 0.05 \\
      & \ldots                  & $8.0\ \micron$ & 0.05 &       & \ldots                  & $8.0\ \micron$ & 0.05 \\
      & \textit{Spitzer}/MIPS   & $24\ \micron$  & 0.05 &       & \textit{Spitzer}/MIPS   & $24\ \micron$  & 0.05 \\
      & \ldots                  & $70\ \micron$  & 0.10 &       & \ldots                  & $70\ \micron$  & 0.10 \\
      & \textit{Herschel}/PACS  & $100\ \micron$ & 0.05 &       & \textit{Herschel}/PACS  & $100\ \micron$ & 0.05 \\
      & \ldots                  & $160\ \micron$ & 0.05 &       & \ldots                  & $160\ \micron$ & 0.05 \\
      & \textit{Herschel}/SPIRE & $250\ \micron$ & 0.15 &       & \textit{Herschel}/SPIRE & $250\ \micron$ & 0.15 \\
\enddata
\tablenotetext{a}{Calibration uncertainty as a fraction of flux: $\sigma_{\rm total} = \sqrt{\sigma^2 + (\sigma_{\rm cal}^{f} f)^2}$}
\end{deluxetable*}

% -----------------------------------------
% 3. SED Fitting
% -----------------------------------------
\section{Spectral Energy Distribution Fitting}
\label{sec:sed}

%% -----------------------------------------
%% 3.1 Stellar Population and Dust
%% -----------------------------------------
\subsection{Stellar Population and Dust Modeling}
\label{sec:sed:stellar}
The host galaxy stellar populations are modeled using piecewise constant (also called ``non-parametric'') SFHs, with constant SFR in each of a number of stellar age bins. For the majority of our sample we use 4 SFH bins, spanning $0$--$10$ Myr, $10$--$100$ Myr, $\rm 100\ Myr$--$\rm 1\ Gyr$, and $5\ {\rm Gyr}$--$t_{\rm age}(z)$, where $t_{\rm age}(z)$ is the age of the Universe at the redshift $z$ of the galaxy. For galaxies with $t_{\rm age}(z) > 5$ Gyr ($z < 1.17$ in our assumed cosmology), we use 5 bins, spanning $0$--$10$ Myr, $10$--$100$ Myr, $\rm 100\ Myr$--$\rm 1\ Gyr$, $1$--$5$ Gyr, and $5\ {\rm Gyr}$--$t_{\rm age}(z)$. Our single age stellar population models are generated with \textsc{P\'egase} \citep{fioc1997,fioc1999}, assuming solar metallicity, $Z = 0.020$, and a \citet{kroupa2001} IMF. These stellar populations also include nebular emission associated with star formation \citep[see section 2.4 of][]{fioc1997}. 

Dust attenuation is modeled with the \citet{noll2009} modified \citet{calzetti2000} attenuation curve, which includes a UV bump and variable UV attenuation curve slope. For the SSA22 sample we fix the UV slope to match the featureless \citet{calzetti2000} law, while for the CDF samples we allow the deviation $\delta$ in the UV slope to vary, as the presence of IR data at the peak of the warm dust emission and the relatively deep optical data allow us to constrain $\delta$.

Dust emission is modeled using the \citet{draine2007} templates. We use the ``restricted'' version of the model set, as recommended by \citet{draine2007}, with the radiation intensity distribution slope fixed at $\alpha = 2$ and the maximum intensity fixed at $U_{\rm max} = 3 \times 10^{5}$. The minimum intensity, $U_{\rm min}$, cannot be well-constrained by our limited IR data for the SSA22 sample, so we (arbitrarily) fix $U_{\rm min} = 10$, finding that this choice also results in acceptable fits for the CDF AGN. The mass fraction of polycyclic aromatic hydrocarbons (PAH) in the dust mixture is also fixed for all of our samples at $q_{\rm PAH} = 0.0047$, which is the minimum for the \ligh\ dust emission implementation, as we do not expect strong PAH emission from the high-redshift galaxies that we are most interested in. We assume energy balance, such that the normalization of the dust emission is set by the UV-optical attenuation of the stellar population and AGN. 

%% -----------------------------------------
%% 3.2 UV-IR AGN Models
%% -----------------------------------------
\subsection{UV-IR Active Galactic Nuclei Modeling}
\label{sec:sed:AGN}

Inspired by the methods adopted in \textsc{CIGALE} \citep{yang2020, yang2022}, we adopt the SKIRTOR UV-IR AGN templates \citep{stalevski2016} in our modeling. These models include a broken power law model for the accretion disk emission, which is reprocessed by a two-phase, clumpy torus of dust surrounding the accretion disk. For the accretion disk component, we use the default SKIRTOR broken power law, where
\begin{equation}
\lambda L_\lambda \propto
	\begin{cases}
		\lambda^{1.2} & 0.001\ \micron \leq \lambda \leq 0.01\ \micron\\
		\lambda^{0} & 0.01\ \micron < \lambda \leq 0.1\ \micron\\
		\lambda^{-0.5} & 0.1\ \micron < \lambda \leq 5\ \micron\\
		\lambda^{-3} & 5\ \micron < \lambda \leq 50 \micron\\
	\end{cases}.
\end{equation}
We use a slice of the model set with the geometry and structure of the torus fixed. We adopt a torus opening angle $40^{\circ}$ (corresponding to a covering factor of $\sin 40^{\circ}\approx0.64$) based on the findings of \citet{stalevski2016}. The amount of dust in the torus is also fixed: in the SKIRTOR models this is controlled by the parameter $\tau_{9.7}$, the edge-on optical depth of the torus at $9.7\ \micron$ (the wavelength of the commonly-seen silicate absorption feature in the spectra of AGN). We adopt $\tau_{9.7} = 7$; we lack the IR data to properly constrain this silicate feature, so we choose the midpoint of the gridded values in the SKIRTOR models. To allow treatment of the viewing angle $i$ as a continuous parameter (the original models are gridded in 10 degree increments), we interpolate between models in $\cos i$-space. 

The UV-optical component of the AGN is attenuated by the same attenuation curve as the stellar population, and we apply energy balance to model heating of the ISM dust by the AGN, such that the power attenuated from the AGN spectrum by ISM dust (integrated over all lines of sight) is added to the luminosity of the dust model. 

Our AGN model does not include line emission, which is a strong and ubiquitous feature in the UV-to-IR spectra of AGN. In early tests we found clear discrepancies for some AGN candidates in the CDF between bands probing the Lyman-$\alpha$ line and our models, where the observed flux was significantly larger than the model was capable of producing, while the neighboring bands probing continuum emission were well-fit. Our stellar population models include Ly$\alpha$, but could not produce the observed fluxes without extreme star formation rates inconsistent with the other observations. We take this to suggest that Ly$\alpha$ emission from the AGN may be contaminating these observations. We note also that X-ray AGN candidates are also often LAEs at redshifts $z\sim3$, including in SSA22, where five out of eight of the \citetalias{lehmer2009a} X-ray AGN were originally discovered as LAEs. Since we do not model AGN line emission, for the CDF samples we exclude bands that contain the Lyman-$\alpha$ line (i.e., $U$-band at $z\sim2$ and ACS F435W at $z\sim3$). 

Some of the strongest AGN narrow lines, including H$\beta$, $\rm [OIII]\lambda5007$, and $\rm [OII]\lambda3727$ fall in the NIR $JHK$ bandpasses at $z=3.09$. One of the SSA22 galaxies, J221732.00+001655.6, has significantly elevated $JHK$ measurements compared to our best-fit models, and in preliminary fits was unacceptably under-fit. Since there is no evidence of a companion galaxy in the F160W image, we speculate that the elevated photometry is due to contamination of the NIR photometry by AGN narrow line emission, and consequently exclude the $J$, $H$, F160W, and $K$ bands from the fit. We note that if strong narrow lines alone caused this issue we might also expect the $i-$band measurement to be larger than the models could produce, due to contributions from the $\rm CIV\lambda1549$ line. We do not see this in our fits, and we cannot conclusively say that narrow lines are the cause. However, we find that the solution we recover without the $J$, $H$, F160W, and $K$ bands is consistent with the preliminary, under-fit solution, though with better sampling and larger uncertainties due to the smaller $\chi^2$. Since the SFH solution for this galaxy is apparently robust to the omission of this photometry, we proceed with the second, better-fit solution.

%% -----------------------------------------
%% 3.3 X-ray Models
%% -----------------------------------------
\subsection{X-ray Modeling}
\label{sec:sed:xray}

The relationship between the intrinsic optical and X-ray luminosities of AGN are typically quantified empirically with a two-point spectral index between the $L_{2500}$ and $L_{\rm 2\ keV}$, the intrinsic luminosity densities at 2500 $\rm \AA$ and 2 keV. There are a number of different calibrations for this relationship, including the $L_{\rm 2500} - \alpha_{ox}$ relationship \citetext{e.g., \citealp{just2007}; as used in \textsc{CIGALE}, \citealp{yang2020, yang2022}} and the \citet{lusso2017} $L_{\rm 2500} - L_{\rm 2\ keV}$ relationship. The scatter around and variations between these empirical relationships were modeled physically in \citet{kubota2018} as variations in the properties of the SMBH, including the SBMH mass $M_{\rm SMBH}$ and the Eddington ratio, $\dot m = \dot M / \dot M_{\rm edd}$. To model the AGN X-ray emission in \ligh, we adopt the \texttt{qsosed} models from \citet{kubota2018}, which were found to reproduce the \citet{lusso2017} $L_{\rm 2500} - L_{\rm 2\ keV}$ relationship throughout the $M_{\rm SMBH} - \dot m$ parameter space \citep[see Figure 7 in][]{kubota2018}. These models, which have previously been applied to X-ray-to-IR SED fitting in \citet{azadi2023}, are a subset of the \texttt{agnsed} family of models, intended to reproduce the soft X-ray excess seen in some AGN spectra with a three-component model including an outer accretion disk, a warm Comptonizing region covering the disk, and a hot Comptonizing region around the central source, which produces the hard X-ray spectrum. The \texttt{qsosed} model fixes the electron temperatures and geometries of the three regions, such that $M_{\rm SMBH}$ and $\dot m$ are the only remaining physical parameters. In \ligh\ we sample both these parameters, and allow them to set the normalization of the entire X-ray-to-IR AGN model by matching the intrinsic $L_{\rm 2500}$ of the SKIRTOR UV-IR AGN model to that of the X-ray model. 

The X-ray emission from X-ray binaries within the stellar population of a galaxy can in some cases rival the X-ray emission from an AGN, though in our samples selected by X-ray luminosity they are expected to have small or negligible contributions to the X-ray spectrum. We model the X-ray binary population as a power law with an exponential cutoff at high energies, with photon index $\Gamma = 1.8$ and cutoff energy $E_{\rm cut} = 100$ keV. We determine the X-ray binary luminosities using $L_{\rm X} / M_{\star}$ relationships based on the stellar-age-dependent parameterizations provided in \citet{gilbertson2022}:

\begin{equation}
	\frac{L^{\rm LMXB}_{\rm X}}{M_{\star}}(\tau) = -1.21(\tau - 9.32)^2 + 29.09 \rm\ erg\ s^{-1} M_{\odot}^{-1}
\end{equation}
and

\begin{equation}
	\frac{L^{\rm HMXB}_{\rm X}}{M_{\star}}(\tau) = -0.24(\tau - 5.23)^2 + 32.54 \rm\ erg\ s^{-1} M_{\odot}^{-1},
\end{equation}
where $\tau$ is the base-10 log of the stellar age in years and $L_{\rm X}$ is measured from rest-frame $2-10$ keV. We thus calculate the HMXB and LMXB luminosity of each stellar age bin of the SFH and sum them to produce the X-ray binary luminosity of a given galaxy.

We model X-ray absorption with the \texttt{tbabs} absorption model \citep{wilms2000}. We use three instances of \texttt{tbabs} to model absorption: two \texttt{tbabs} in the rest-frame of the galaxy to model the absorption of the AGN and the X-ray binary population, and one in the observed-frame to account for the absorption by the Milky Way, where the Galactic HI column density along the line of sight (retrieved using the \texttt{CIAO} \texttt{colden} procedure) is held constant. Only the HI column density in the nuclear regions is a free parameter in our fits; the HI column density in the galaxy (modifying only the stellar population absorption) is derived from the $V-$band attenuation, $A_V$ (computed in turn from $\tau_V$) assuming that

\begin{equation}
	N_H / A_V = 2.24 \times 10^{21}\ {\rm cm^{-2}\ mag^{-1}},
\end{equation}
chosen as an average of observations of the Milky Way $N_H-A_V$ relationship \citep{predehl1995, nowak2012}. We note that the results for our samples, which we selected by $L_X$ to include only AGN-dominated sources, are insensitive to the assumed level of absorption for the X-ray binary population.
%In \texttt{XSpec} terms, then, our full X-ray model is \texttt{tbabs * (tbabs * qsosed + tbabs * plaw)}

Our X-ray fitting procedure in \ligh\ can use either fluxes, which are widely available in high-level X-ray source catalogs but are dependent on an assumed spectral model, or instrumental counts, which are model-independent but require ancillary products to use in fitting. In the former case, the user need only supply fluxes, uncertainties, and bandpasses. Bandpasses are assumed to have uniform sensitivity across their width. In the latter case, we require a user to supply the net counts in each bandpass, the bandpasses, the exposure time, and the auxiliary response function (ARF), which describes the energy-dependent effective area of the detector. The models are converted to photon flux density, multiplied by the ARF, and integrated over the bandpass to get the counts predicted by the model in each bandpass. In this work, we use instrumental counts for our X-ray photometry.

For the CDF-N and CDF-S, we use the ARF nearest the aimpoint of the field and account for its off-axis variation by using the vignetting-corrected exposure time from the published CDF point source catalogs. In SSA22, we generate an ARF appropriate for each source position, based on our \textsc{AE} runs. Once we have calculated the model counts, we calculate the X-ray model contribution to $\chi^2$ as 

\begin{equation}
	\chi^2_{\rm Xray} = \sum_{i = 1}^{n} \frac{(N^{mod}_i - N_i)^2}{\sigma N_i^2},
\end{equation}
where $N_i$ is the net counts in bandpass $i$. For the SSA22 data, we use

\begin{equation}
	\sigma N_i^2 = \sigma S_i^2 + (\frac{A_{S}}{A_{B}})^2 \sigma B_i^2,
\end{equation}
where $S_i$ and $B_i$ are the source and background counts in bandpass $i$, with $A_{S}$ and $A_{B}$ the respective areas of the source and background regions. We use the upper error of the \citet{gehrels1986} approximation for the uncertainties:

\begin{equation}
	\sigma x = 1 + \sqrt{3/4 + x}.
\end{equation}
For the far deeper CDF data, we assume the source counts dominate over the background counts in both bandpasses, and let

\begin{equation}
	\sigma N_i = 1 + \sqrt{3/4 + N_i}.
\end{equation}

%% -----------------------------------------
%% 3.4 Model Fitting
%% -----------------------------------------
\subsection{Fitting}
\label{sec:sed:fitting}

The parameters outlined above and their allowed ranges for our SED fits are summarized in \autoref{tab:SEDParams}. We assumed uniform priors on all free parameters. Upper limits are handled in our fitting procedure by setting the observation in the corresponding band to 0, and setting the uncertainty to the 1$\sigma$ limiting flux.\footnote{This is only an approximation of the statistically correct handling of upper limits, albeit a widely-used one that produces the expected behavior for upper limits (see, e.g., \autoref{fig:SSA22_fits_b}). See Appendix A of \citet{sawicki2012} for a discussion of the model likelihood in the case of nondetections.} 

To fit the SEDs of our samples, we made use of an implementation of the \citet{goodman2010} affine-invariant MCMC algorithm within \ligh. We use an ensemble of 75 MCMC samplers, which we run for 40000 steps. We use the ``stretch move'' as described in \citet{goodman2010}. To achieve acceptance fractions $>20\%$, we change the proposal distribution width parameter $a$ to $1.8$ (the original authors recommend using $a=2$). For our fits, we find that the integrated autocorrelation times of the parameters can be quite long, approaching $\sim 1000$ steps for some parameters \citetext{see, e.g., \citealp{goodman2010} and \citealp{foremanmackey2013} for discussions of the autocorrelation time in MCMC applications}. 

To produce the final chains, from which we sample the best-fit parameters and uncertainties, we first discard a number of burn-in steps from the beginning of each chain equal to twice the longest autocorrelation time of any parameter of the chain. We then calculate the autocorrelation time again for the new chain minus burn-in, thin the remaining length of the chains by a factor of half the new longest autocorrelation time, and stack the chains of all the samplers. This results in a final chain consisting of independent samples. We take the last 1000 of these samples for each galaxy as the sampled posterior distribution. 

When plotting the SEDs, we show the best fit and the full range of the best 68\% of models (in terms of $\chi^2$) drawn from the sampled posterior as the uncertainty in the fit. When quoting individual parameter values and star formation histories, we report the median values from the marginalized posterior distributions, with the 16th to 84th percentile ranges as the uncertainties. We make this distinction to better show the connection between the observational coverage, data uncertainties, and the uncertainties on the fit when plotting the SED.

We use posterior predictive checks (PPC) \citetext{\citealp{rubin1984,gelman1996}; see also section 5.1 of \citealp{chevallard2016} for an example of PPC used in SED fitting} to assess the suitability of the models used in our fits. Of the SSA22 sample, two galaxies, J221736.54+001622.6 ($p = 0.013$) and J221716.16+001745.8 ($p = 0.000$) have $p-$values that suggest poor fits, though we find that these are likely data-driven rather than model-driven.

While the $p-$value for J221736.54+001622.6 is not extreme (i.e. it is $>0.01$), it does suggest under-fitting. Upon inspection, it appears that the under-fitting is due entirely to the $u^*$ measurement. The other galaxies in the SSA22 sample have $u^*$ measurements consistent with either non-detection or a steep fall-off in flux, as expected for $z = 3.1$ galaxies; J221736.54+001622.6 has the largest $u^*$ flux of the SSA22 galaxies (a factor of $\sim 10$ greater than the next-largest). This suggests that J221736.54+001622.6 is an outlier; given the good agreement between the data and our model for the remainder of the bands, we treat our model as appropriate for this galaxy otherwise.

The fit to J221716.16+001745.8 is significantly ruled out according to the PPC, with $p < 0.001$. Upon inspection, this appears to be driven by the ALMA data: the ALMA Band 6 (1.1 mm) detection is significantly larger than the Band 7 (870 \micron) upper limit, suggesting a flat spectral slope that the Rayleigh-Jeans tail of our dust model cannot reproduce. Notably, this source is located close ($<2''$) to an offset ALMA 870 \micron\ detection not originally associated with this source. However, the offset of the 870 \micron\ source is in the direction of the synthesized beam's major axis, suggesting that it could be a plausible counterpart. We re-fit with the flux of the offset detection in place of the upper limit, finding a much improved fit ($p = 0.164$), and the same SFR and SFH shape as we recover when fitting with the ALMA Band 7 upper limit. See \autoref{app:AGN7_alternative} for details on the alternative SED fit to this AGN.

To establish the degree to which we can constrain the stellar population properties of the SSA22 AGN hosts, we also fit their SEDs with an AGN-only model (i.e., with all SFH coefficients set to 0). Details of these AGN-only SED fits and comparisons to the model with a stellar population can be found in \autoref{app:SSA22_pure_AGN}. For all but two of the SSA22 AGN, the PPC rules out the AGN-only model with $p < 0.001$. Our model-comparison tests indicate that the AGN-only model is statistically preferred for both of these AGN, J221720.24+002019.3 and J221759.23+001529.7. Given that the data (under our models) are consistent with zero star formation, in what follows we treat the stellar population properties from our original SED fits as upper limits for these two AGN.

As for the CDF sample, the majority of galaxies are well-fit with $p > 0.05$, though we find 35/151 = $23\%$ (24/151 = $16\%$) with $p < 0.05$ ($0.01$). This under-fitting may be model-driven (i.e. the model is not suitable for the data) or data-driven (if, e.g. the data are blended, mismatched, or the uncertainties are overly-constraining). We have taken steps to mitigate blending of the IR data, and used the carefully cross-matched \citet{barro2019} catalogs to attempt to mitigate mismatches. We also added calibration uncertainties to the data to avoid overly-constrictive uncertainties. However, we note that our choice of model uncertainty ($10\%$ in each band) is arbitrary; we may need larger values to appropriately model uncertainties about, e.g. the power law slopes of the AGN accretion disk model, the optical depth of the AGN torus, or the values of any of the other fixed model parameters. To test the effect of model uncertainty on the distribution of $p-$values, we re-fit the CDF samples with a model uncertainty of $12\%$ in each band. This decreased the number of galaxies with $p < 0.01$ to 11/151 ($7\%$). We note also that across the sample, the under-fitting does not appear to be systematically related to any particular component of the model. Given the decrease in under-fitting with a modest and arbitrarily chosen increase in the model uncertainty, and the lack of systematics in the under-fitting, we treat the models as suitable for the CDF sample.

%%% -----------------------------------------
%%% Table: SED fit parameters
%%% -----------------------------------------
\begin{deluxetable*}{l l l r}[ht]
\tablecolumns{4}
\tablecaption{\label{tab:SEDParams}Parameters and assumptions for SED fits.}
\tablehead{\colhead{Model Component} & \colhead{Parameter} & \colhead{Description} & \colhead{Value/Range\tablenotemark{a}}}
\startdata
Stellar Population & $\{\psi_i\}_{i=1}^{n}\tablenotemark{b}$              & Star Formation History coefficients in solar masses per year  & $[0.0, \infty)$ \\
Dust Attenuation   & $\tau_{V,\rm Diff}$                 & Optical depth of diffuse dust                                 & $[0.0, 10.0]$ \\
				   & $\delta$                            & UV attenuation curve power-law slope deviation from \citet{calzetti2000} law & $[-4.0, 0.3]$\tablenotemark{c} \\
Dust Emission      & $\alpha$                            & Power law slope of intensity distribution                     & 2.0 \\
                   & $U_{\rm min}$                       & Intensity distribution minimum                                & 10  \\
                   & $U_{\rm max}$                       & Intensity distribution maximum 		                         & $3\times10^5$ \\
                   & $q_{\rm PAH}$                       & Mass fraction of PAHs in dust mixture                         & 0.0047 \\
                   & $\gamma$                            & Mass fraction of dust exposed to intensity distribution       & $[0.0, 1.0]$ \\
%AGN Emission       & $\log(L_{\rm AGN}/ \rm L_\odot)$    & UV-IR AGN normalization                                      & $[0.0, 20.0]$ \\ 
AGN Emission	   & $\tau_{9.7\rm\ \micron}$            & Edge-on optical depth of AGN torus at $9.7\ \rm\micron$       & 7.0 \\
%				   & CF\tablenotemark{c}                 & Torus covering factor                                         & 0.64 \\
				   & $\cos i$                            & Cosine of AGN torus inclination to the line of sight          & $[0.0, 1.0]$ \\
X-ray Emission     & $n_{H} / 10^{20}\rm\ cm^{-2}$       & Neutral Hydrogen column density along the line of sight       & $[10^{-4}, 10^{5}]$\\
				   & $M_{\rm SMBH}$                      & SMBH mass parameter in solar masses					         & $[10^{5}, 10^{10}]$\\
				   & $\log \dot m$						 & $\log_{10}$ of SMBH accretion rate, normalized by the Eddington rate & $[-1.5, 0.3]$
\\
\enddata
\tablenotetext{a}{For free parameters, the allowed ranges are given in brackets. Priors are uniform on all free parameters.}
\tablenotetext{b}{Galaxies with $z < 1.17$ have 5 SFH coefficients; galaxies with $z \ge 1.17$ (including the entire SSA22 sample) have 4.}
\tablenotetext{c}{$\delta$ is only a free parameter for the CDF sample fits; it is fixed at 0 for the SSA22 sample.}
%\tablenotetext{b}{Where we lack sufficient IR data, we fit with $U_{\rm min}$ fixed at 10.}
%\tablenotetext{c}{The covering factor is the sine of the opening angle of the dusty cone of the torus; the adopted value corresponds to an opening angle of $40^{\circ}$.}
\end{deluxetable*}

% -----------------------------------------
% 4. Results
% -----------------------------------------
\section{Results}
\label{sec:results}

\subsection{\citetalias{lehmer2009a} Active Galact Nuclei}

\begin{figure*}[p]
\centering
\includegraphics[width=\textwidth]{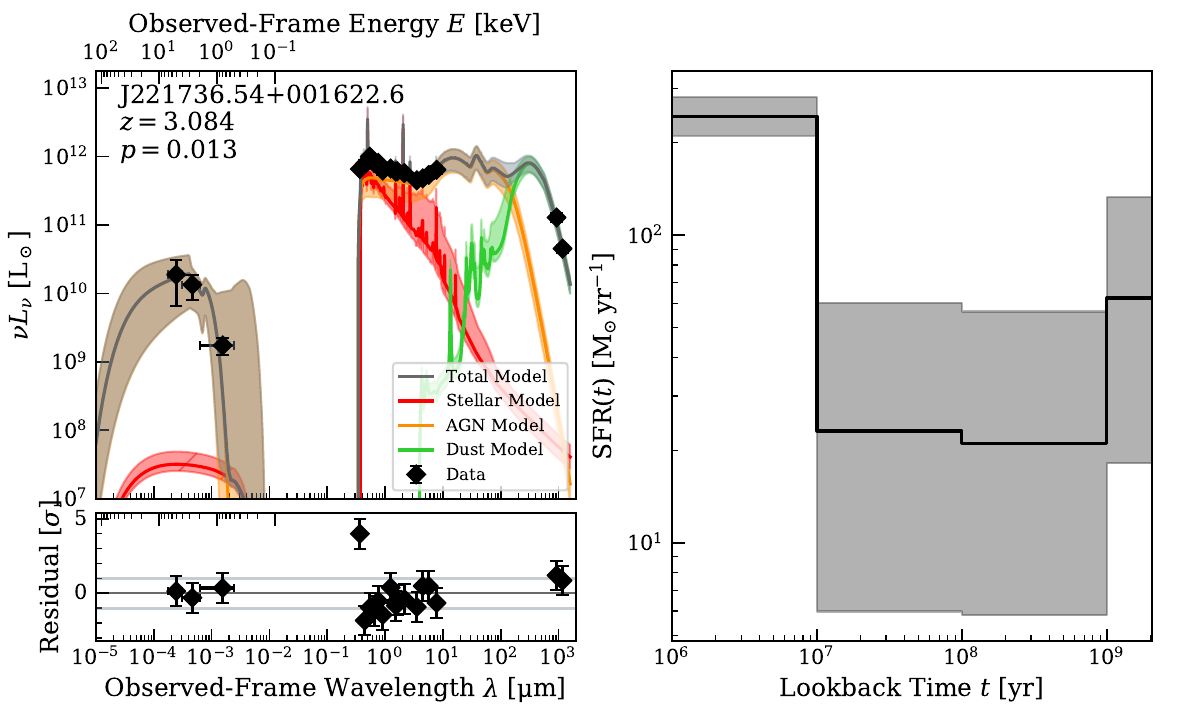}
\includegraphics[width=\textwidth]{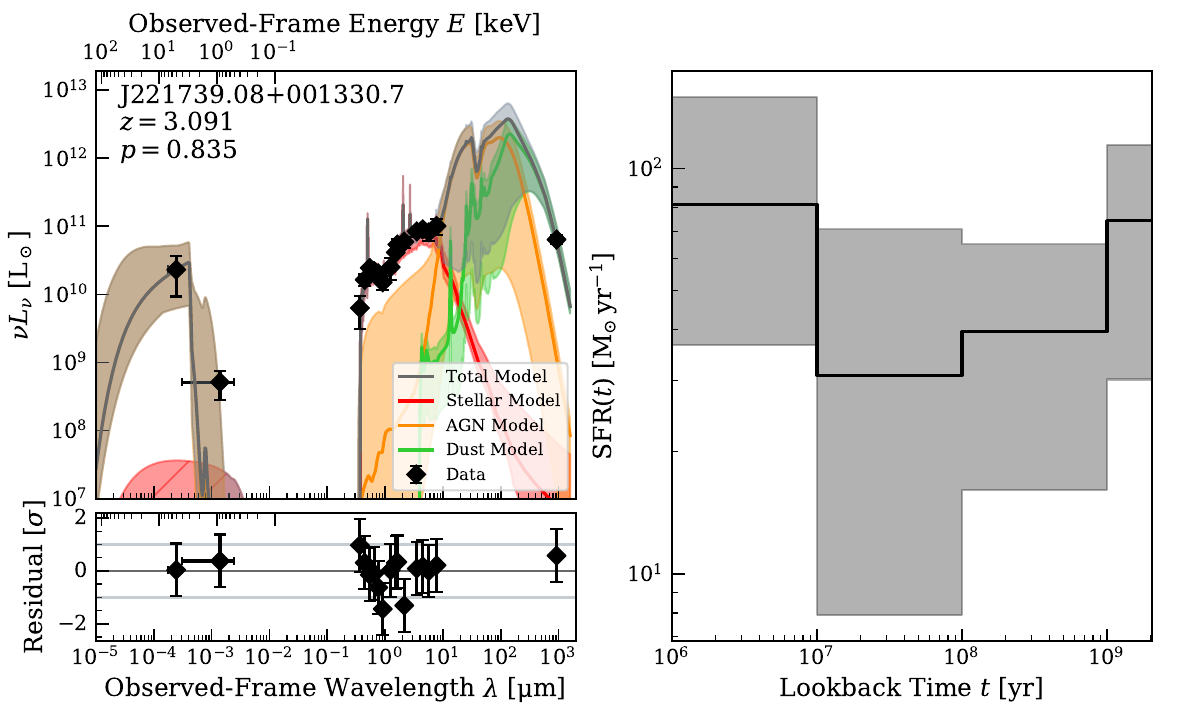}
\caption{\label{fig:SSA22_fits_a}\textit{Left panels:} The best-fit SED model for each of the 8 \citetalias{lehmer2009a} X-ray AGN in the SSA22 is shown with its components. The shaded regions of the SED plot indicate the full range of the best-fitting 68\% of models. Upper limits are shown at $3\sigma$. The data $-$ model residuals from the best-fit model are shown below, in units of $\sigma$. The $p-$value in the annotation is computed with a posterior predictive check (see \autoref{sec:sed:fitting}). \textit{Right panels:} We show the sampled posterior SFH: for galaxies with constrained SFH, the dark line indicates the median of the posterior and the shaded region shows the 16th to 84th percentile range. For J221720.24+002019.3 and J221759.23+001529.7 the dark line indicates the 99th percentile upper limit of the posterior.}
\end{figure*}

In \autoref{tab:SSA22_properties} we summarize the SED-fitting derived physical properties of the SSA22 AGN sample. Best-fit SED models and sampled posterior SFHs for the 8 \citetalias{lehmer2009a} AGN are shown in \autoref{fig:SSA22_fits_a}, with ADF22A1 shown in \autoref{fig:ADF22A1_sed} (see \autoref{sec:results:adf22a1}). The stellar masses and star formation rates for the SSA22 and CDF samples are shown in \autoref{fig:main_sequence}. We find stellar masses ranging from $10^{10.5}-10^{11.0}\ {\rm M_{\odot}}$ and black hole masses ranging from $10^{8.3}-10^{8.6}\ {\rm M_{\odot}}$ for the protocluster AGN, though our measurements of the black hole mass may be complicated by the level of circumnuclear obscuration in the sample. Of the galaxies in the SSA22 sample for which we can successfully constrain the star formation history, half (3/6) are classified as sub-main sequence ($-1.3 < \log {\rm SFR/SFR_{MS}} \leq -0.4$) or high quiescent ($\log {\rm SFR/SFR_{MS}} \leq -1.3$) according to the criteria and redshift-dependent main sequence defined by \citet{aird2019}. In particular, J221709.60+001800.1 is identified as a plausible quiescent galaxy, with ${\rm SFR} = 6.80^{+5.20}_{-3.24}\ {\rm M_{\odot}\ yr^{-1}}$ and $\log{\rm SFR/SFR_{MS}} = -1.38^{+0.26}_{-0.29}$. The galaxies consistent with the main sequence (J221735.84+001559.1, J221716.16+001745.8, and J221732.00+001655.6) are typically ALMA-detected, dusty galaxies, with $\tau_V = 1.4 - 2.3$. Of the galaxies with $\log {\rm SFR/SFR_{MS}} \leq -0.4$, only J221736.54+001622.6 is ALMA-detected. Our estimates of the SFR for the ALMA Band 7 detected galaxies are compatible with the SFR estimates from \citet{alexander2016}, which were based on a \citet{kennicutt1998} scaling relationship corrected to a \citet{chabrier2003} IMF. The \citet{alexander2016} estimates are systematically larger by a factor of $\sim 2$, but this is to be expected when comparing SFR estimates from scaling relations and SED fitting; our fits to these galaxies typically include significant components in the SFH from stars older than 100 Myr, which also heat dust and contribute to the measured $L_{IR}$. As \citet{alexander2016} estimated $L_{IR}$ from the flux at 870 \micron, well beyond the peak of the AGN IR emission, their SFR estimates are unlikely to be contaminated by AGN emission.

The majority (6/8) of the fits to the SSA22 AGN sample are consistent with heavily obscured AGN growth ($N_H \geq 5 \times 10^{23}\ \rm cm^{-2}$), and J221759.23+001529.7 is moderately obscured ($N_H = 1$--$5 \times 10^{23}\ {\rm cm^{-2}}$) with $N_H = 4.37^{+2.16}_{-1.92} \times 10^{23}\ {\rm cm^{-2}}$. Three of the heavily obscured AGN are Compton-thick (CT) candidates, with $N_H \geq 1 \times 10^{24}\ {\rm cm^{-2}}$ (J221709.60+001800.1 is also consistent with Compton-thickness, with $N_H = 9.98^{+45.97}_{-6.17} \times 10^{23}$), which makes the estimates of their black hole masses and particularly accretion rates more uncertain, as our X-ray model does not include line emission or reflection, which make important contributions to the X-ray spectra of CT sources. The level of obscuration we measure from our SED fits is consistent with the (typically hard) spectral indices estimated by \citet{lehmer2009b}: notably, they found $\Gamma < 0.42$ for J221739.08+001330.7 and $\Gamma = 0.16$ for J221732.00+001655.6, both of which we find to be CT candidates.

Only J221720.24+002019.3 is consistent with being lightly obscured or unobscured, with $N_H = 4.87^{+6.15}_{-3.21} \times 10^{22}\ {\rm cm^{-2}}$. J221720.24+002019.3 also has low UV-optical attenuation ($\tau_V = 0.2 \pm 0.1$) and our proxy for circumnuclear UV-optical obscuration, ${\rm cos} i$, is consistent with an unobscured view of the SMBH, suggesting that this source is a so-called ``blue'' quasar. This source, along with J221736.54+001622.6 and J221716.16+001745.8 are presented in the literature as Type 1 AGN with broad lines in their rest-frame optical spectra \citetext{J221720.24+002019.3 in \citealp{yamada2012}, J221736.54+001622.6 in \citealp{steidel2003}, and J221716.16+001745.8 in \citealp{saez2015}}. We find that AGN continuum emission makes strong contributions to the optical SED fits for J221720.24+002019.3 and J221736.54+001622.6, but not in the case of J221716.16+001745.8. This may suggest additional dust obscuration along the line of sight in J221716.16+001745.8, beyond the circumnuclear region. The broad line emission from unobscured AGN is one of the primary obstacles to extracting detailed star formation histories; we note that while we fit an AGN continuum component to the rest-frame UV-optical emission, we are unable to account for broad lines, and so the SFH we present here are naturally more uncertain. The ALMA Band 7 detections for J221736.54+001622.6 and J221716.16+001745.8 allow us to place good constraints on their recent star formation, though the constraints on the oldest two stellar age bins are weaker, especially in the case of J221736.54+001622.6, where the strength of the AGN continuum emission dominates over the stellar population at rest-frame wavelengths $\gtrsim 0.2\ \micron$. In the case of J221720.24+002019.3, the overall shape and normalization of the SFH are not well constrained by our original SED fit, though the non-detection in both ALMA bands suggests low star-formation activity. We find a SFR of $\leq 53.83\ {\rm M_{\odot}}$ (a factor of 3 lower than the upper limit SFR computed from the ALMA limiting flux in \citet{alexander2016}), and the upper limit on $\log{\rm SFR/SFR_{\rm MS}}$ is $-0.31$, suggesting that this galaxy is at most a factor of $\approx 2$ below the location of the star-forming main sequence at $z=3.09$ and is likely a sub-MS galaxy. The low SFR is also supported by optical spectroscopy: the SDSS spectrum\footnote{Retrieved from \url{http://skyserver.sdss.org/dr17}} has a clear, strong 4000 \AA\ break, though at this redshift the 4000 \AA\ break is at the edge of the SDSS sensitivity. The low SFR limits we recover for this source, in combination with the high X-ray luminosity and low obscuration (as measured by both $N_H$ and $\cos i$), place it in an interesting position in the ``story'' of AGN growth in the protocluster, as a possible endpoint of AGN evolution, where star formation has stopped and the AGN has blown away obscuring material.

%%% -----------------------------------------
%%% Figure: AGN SEDs (a)
%%% -----------------------------------------

%%% -----------------------------------------
%%% Figure: AGN SEDs (b)
%%% -----------------------------------------
\begin{figure*}[p]
\centering
\figurenum{4}
\includegraphics[width=\textwidth]{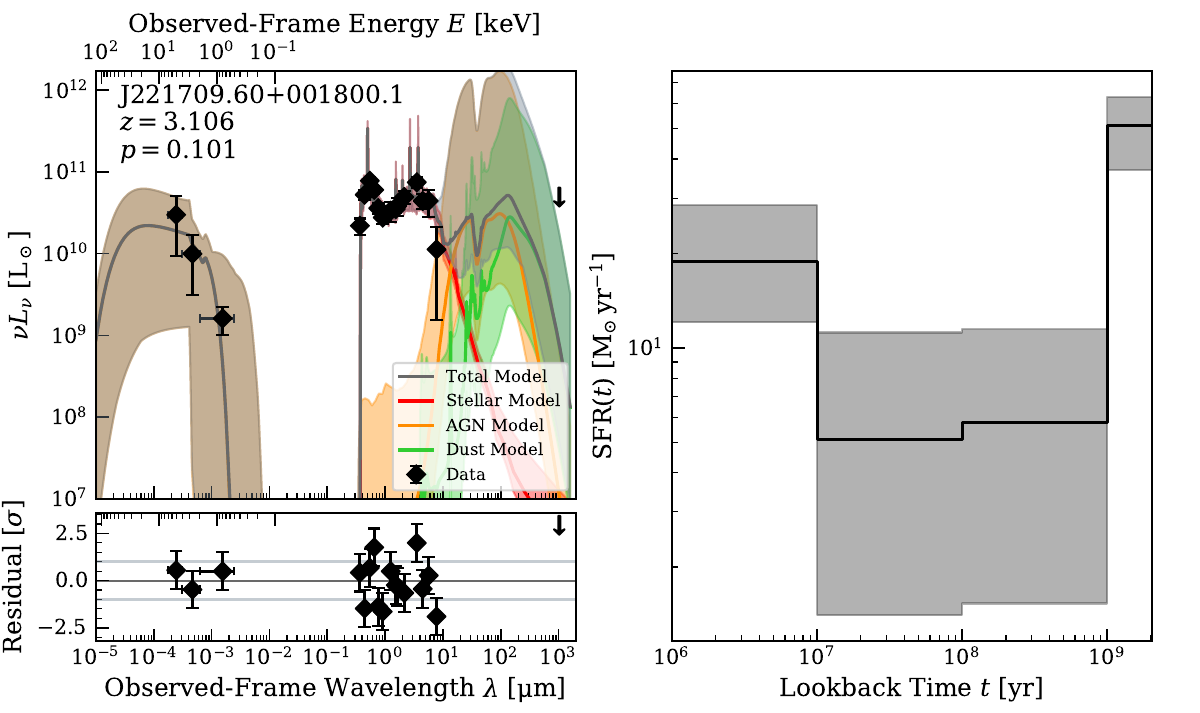}
\includegraphics[width=\textwidth]{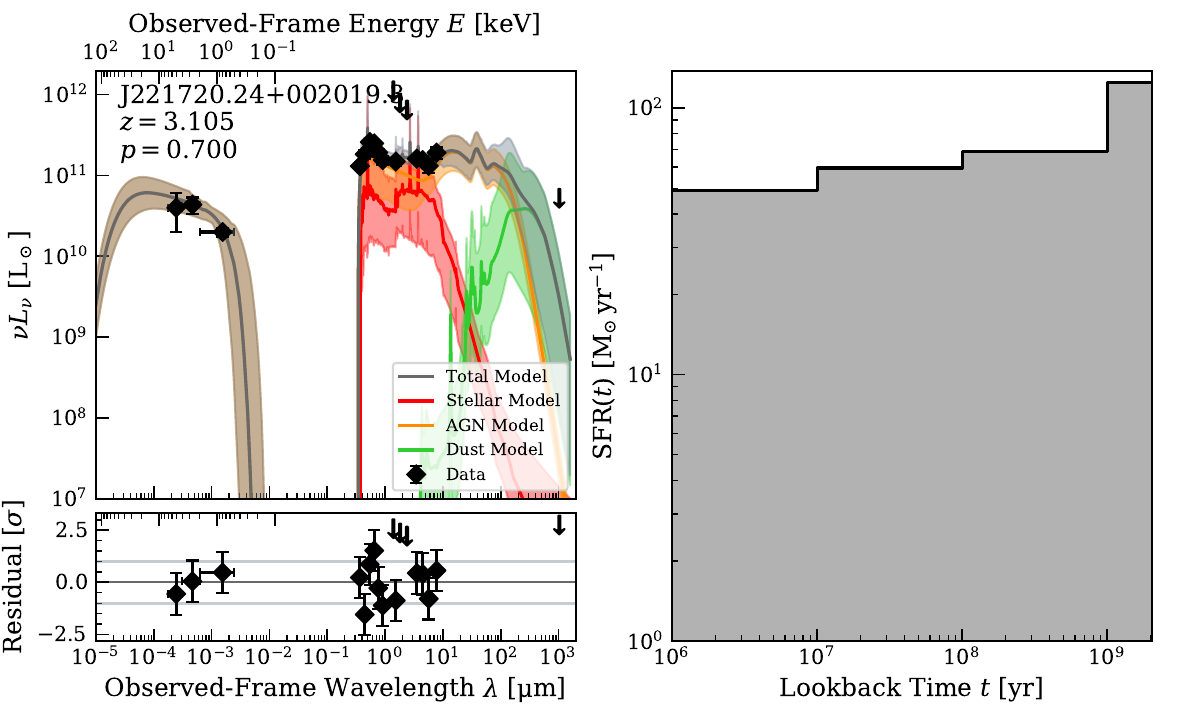}
\caption{\label{fig:SSA22_fits_b}\textit{Continues.}}
\end{figure*}

%%% -----------------------------------------
%%% Figure: AGN SEDs (c)
%%% -----------------------------------------
\begin{figure*}[p]
\centering
\figurenum{4}
\includegraphics[width=\textwidth]{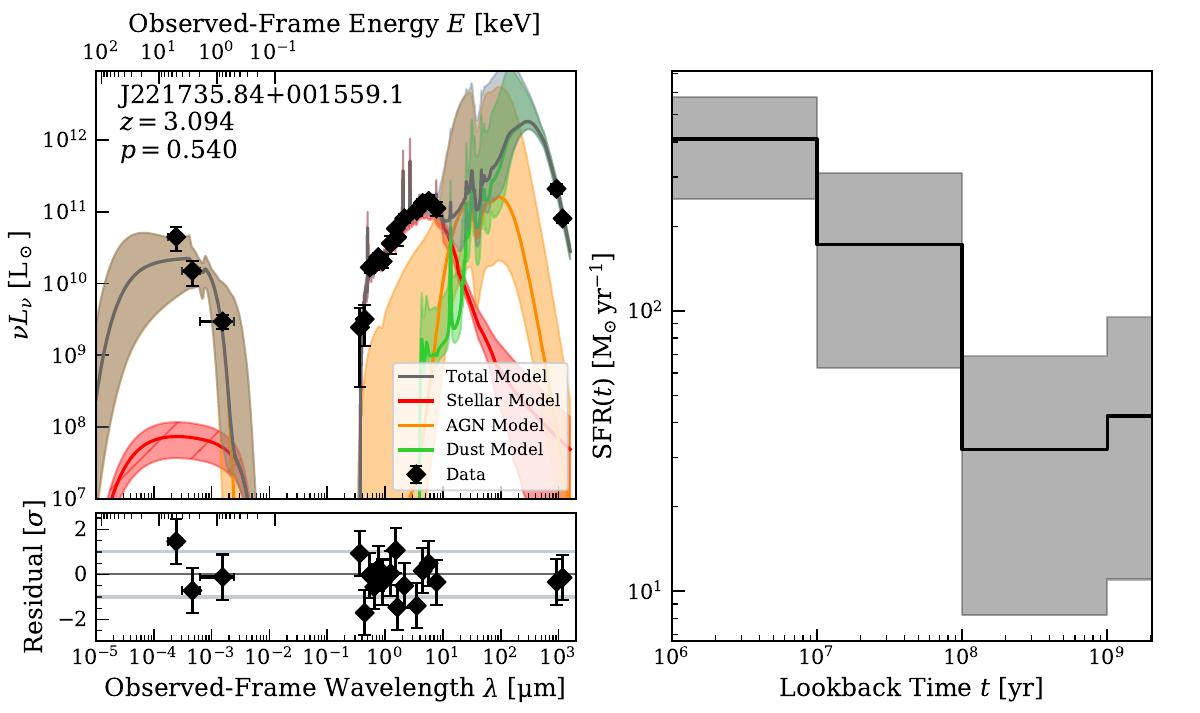}
\includegraphics[width=\textwidth]{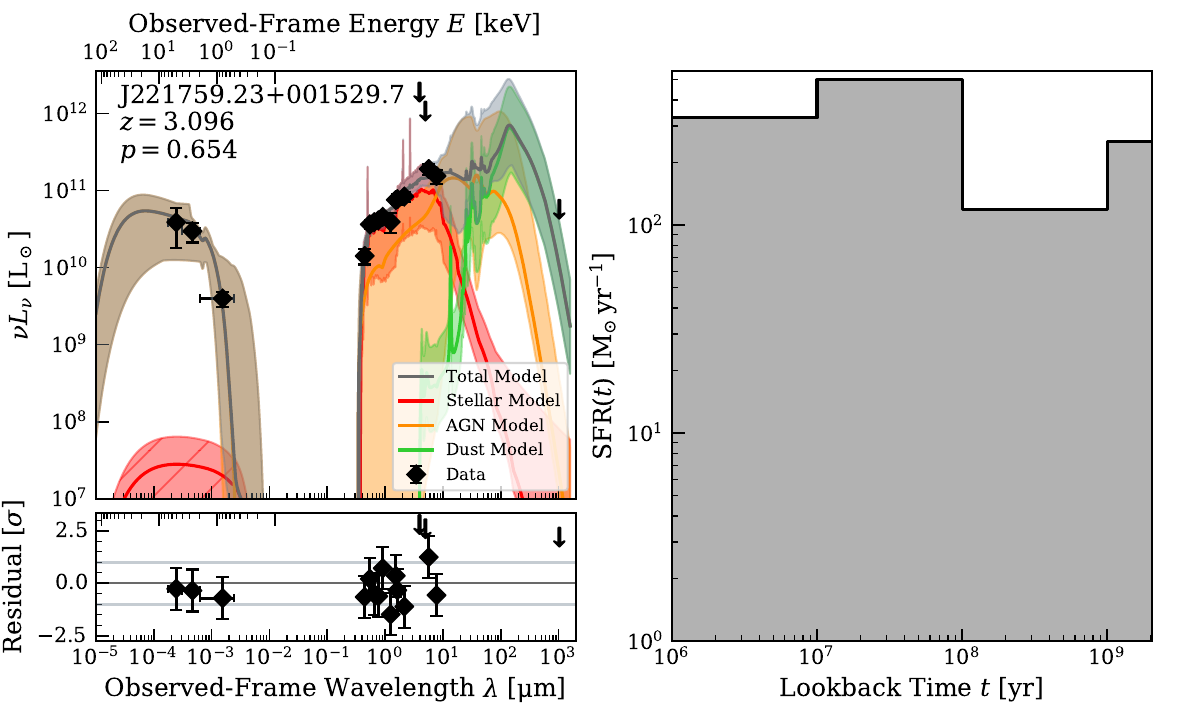}
\caption{\label{fig:SSA22_fits_c}\textit{Continues.}}
\end{figure*}

%%% -----------------------------------------
%%% Figure: AGN SEDs (d)
%%% -----------------------------------------
\begin{figure*}
\centering
\figurenum{4}
\includegraphics[width=\textwidth]{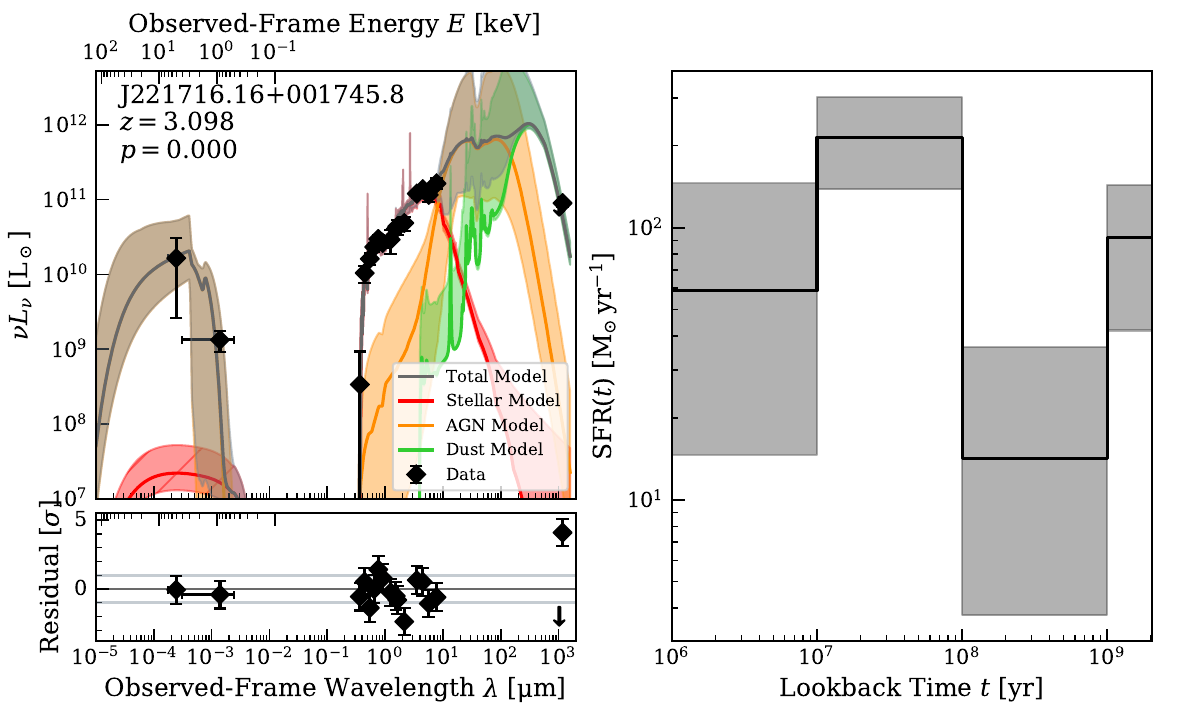}
\includegraphics[width=\textwidth]{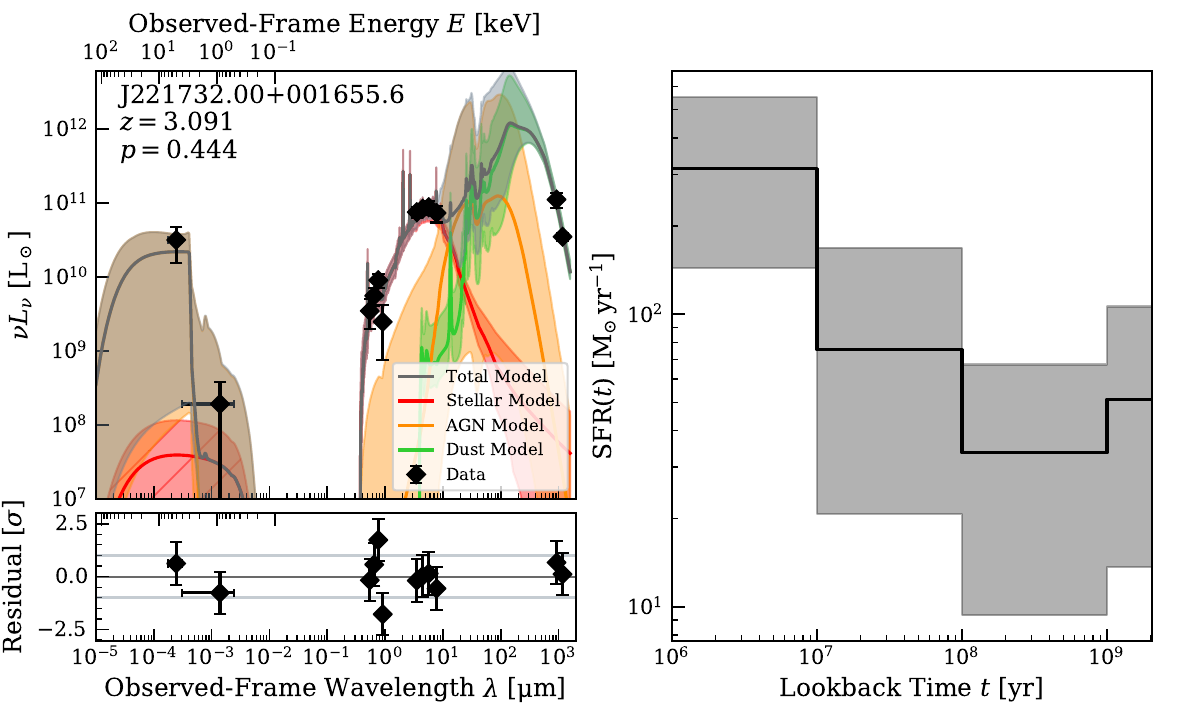}
\caption{\label{fig:SSA22_fits_d}\textit{Continues.}}
\end{figure*}

% -----------------------------------------
% 4.X ADF22A1
% -----------------------------------------
\subsection{ADF22A1}
\label{sec:results:adf22a1}

\begin{deluxetable*}{l r r r r r r r r}
\tablecolumns{9}
\tablecaption{\label{tab:SSA22_properties}SED-fit derived properties for the SSA22 sample.}
\tablehead{\colhead{ID} & \colhead{$\log M_{\star}$\tablenotemark{a}} & \colhead{SFR\tablenotemark{b}} & \colhead{${\rm log SFR/SFR_{MS}}$\tablenotemark{c}} & \colhead{$\tau_V$\tablenotemark{d}} & \colhead{$\cos i$\tablenotemark{e}} & \colhead{$\log M_{\rm SMBH}$\tablenotemark{f}} & \colhead{$\log \dot m$\tablenotemark{g}} & \colhead{$N_H$\tablenotemark{h}} \\ \colhead{} & \colhead{$(\rm M_\odot)$} & \colhead{($\rm M_\odot\ yr^{-1}$)} & \colhead{(dex)} & \colhead{} & \colhead{} & \colhead{$(\rm M_\odot)$} & \colhead{} & \colhead{($10^{22}\ \rm cm^{-2}$)}}
\startdata
J221736.54+001622.6&    $10.8^{+0.2}_{-0.3}$&    $45.2^{+31.6}_{-13.8}$&    $-0.72^{+0.33}_{-0.25}$&    $0.3^{+0.0}_{-0.0}$&    $0.80^{+0.10}_{-0.04}$&    $8.3^{+0.1}_{-0.1}$&    $-0.5^{+0.3}_{-0.2}$&    $59.0^{+35.8}_{-23.9}$\\
{\bf J221739.08+001330.7\tablenotemark{*}}&    $\bf 10.9^{+0.1}_{-0.1}$&    $\bf 36.8^{+37.7}_{-21.1}$&    $\bf -0.91^{+0.37}_{-0.36}$&    $\bf 1.2^{+0.2}_{-0.3}$&    $\bf 0.28^{+0.17}_{-0.18}$&    $\bf 8.6^{+0.1}_{-0.2}$&    $\bf -0.3^{+0.3}_{-0.4}$&    $\bf 572^{+299}_{-268}$\\
J221709.60+001800.1&    $10.5^{+0.1}_{-0.1}$&    $6.80^{+5.20}_{-3.24}$&    $-1.38^{+0.26}_{-0.29}$&    $0.2^{+0.1}_{-0.1}$&    $0.20^{+0.21}_{-0.14}$&    $8.2^{+0.2}_{-0.3}$&    $-1.0^{+0.5}_{-0.4}$&    $99.8^{+460.0}_{-61.7}$\\
J221720.24+002019.3&    $\leq10.9$\tablenotemark{$\dag$}&    $\leq53.8$&    $\leq-0.31$&    $0.2^{+0.1}_{-0.1}$&    $0.82^{+0.09}_{-0.06}$&    $8.6^{+0.1}_{-0.1}$&    $-1.4^{+0.1}_{-0.0}$&    $4.87^{+6.15}_{-3.21}$\\
J221735.84+001559.1&    $10.8^{+0.1}_{-0.2}$&    $198^{+111}_{-90}$&    $-0.13^{+0.28}_{-0.35}$&    $1.9^{+0.1}_{-0.1}$&    $0.22^{+0.23}_{-0.16}$&    $8.5^{+0.2}_{-0.2}$&    $-0.4^{+0.5}_{-0.7}$&    $52.6^{+32.2}_{-23.5}$\\
J221759.23+001529.7&    $\leq11.3$&    $\leq455$&    $\leq0.34$&    $1.2^{+0.1}_{-0.1}$&    $0.40^{+0.29}_{-0.26}$&    $8.5^{+0.1}_{-0.1}$&    $-1.2^{+0.4}_{-0.2}$&    $43.70^{+21.60}_{-19.10}$\\
{\bf J221716.16+001745.8}&    $\bf 10.9^{+0.1}_{-0.1}$&    $\bf 201^{+76}_{-69}$&    $\bf -0.20^{+0.23}_{-0.24}$&    $\bf 1.4^{+0.1}_{-0.1}$&    $\bf 0.33^{+0.15}_{-0.15}$&    $\bf 8.7^{+0.1}_{-0.2}$&    $\bf -0.1^{+0.2}_{-0.3}$&    $\bf 425^{+375}_{-238}$\\
{\bf J221732.00+001655.6}&    $\bf 10.8^{+0.1}_{-0.2}$&    $\bf 105^{+79}_{-49}$&    $\bf -0.39^{+0.32}_{-0.36}$&    $\bf 2.3^{+0.2}_{-0.3}$&    $\bf 0.22^{+0.23}_{-0.15}$&    $\bf 8.3^{+0.2}_{-0.4}$&    $\bf -0.5^{+0.5}_{-0.7}$&    $\bf 622^{+263}_{-275}$\\
{\bf ADF22A1}&    $\bf 11.3^{+0.2}_{-0.2}$&    $\bf 524^{+276}_{-258}$&    $\bf -0.05^{+0.30}_{-0.40}$&    $\bf 2.9^{+0.1}_{-0.1}$&    $\bf 0.33^{+0.21}_{-0.23}$&    $\bf 8.6^{+0.2}_{-0.2}$&    $\bf -0.8^{+0.3}_{-0.2}$&    $\bf 519^{+324}_{-313}$\\
\enddata
\tablenotetext{a}{Total stellar mass, integrated over the SFH.}
\tablenotetext{b}{Star formation rate, averaged over the last 100 Myr of the SFH.}
\tablenotetext{c}{Ratio of SFR to the main sequence SFR of \citet{aird2019}.}
\tablenotetext{d}{$V-$band optical depth of diffuse dust.}
\tablenotetext{e}{Cosine of the inclination from the AGN axis to the line of sight.}
\tablenotetext{f}{Supermassive black hole mass from \texttt{qsosed} model.}
\tablenotetext{g}{Eddington ratio from \texttt{qsosed} model.}
\tablenotetext{h}{Neutral Hydrogen column density along the line of sight.}
\tablenotetext{*}{Compton-thick (CT) candidates with $n_H \geq 1\times10^{24}\rm\ cm^{-2}$ are highlighted in bold.}
\tablenotetext{$\dag$}{Upper limits for J221720.24+002019.3 and J221759.23+001529.7 are given at the 99th percentile of the posterior.}
\end{deluxetable*}

%%% -----------------------------------------
%%% Figure: ADF22A1 SED
%%% -----------------------------------------
\begin{figure*}[ht]
\centering
\includegraphics[width=\textwidth]{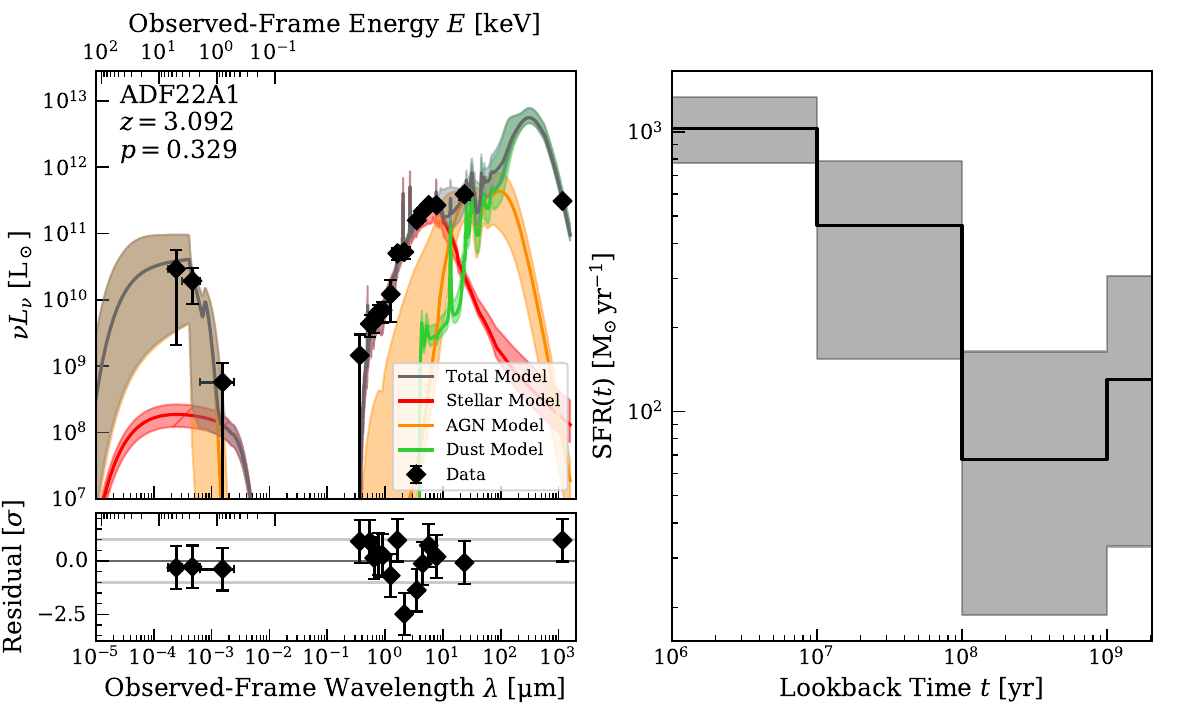}
\caption{\label{fig:ADF22A1_sed}\textit{Left panel:} The best fit SED model for ADF22A1 is shown along with its components. The shaded regions of the SED plot indicate the full range of the best-fitting 68\% of models. The lower panel shows the data $-$ model residuals from the best-fit model, in units of $\sigma$. The $p-$value in the annotation is computed with a posterior predictive check (see \autoref{sec:sed:fitting}). \textit{Right panel:} The sampled posterior distribution of the SFH is shown with the dark line indicating the median of the posterior, and the shaded region showing the 16th to 84th percentile range. The SED and SFH of ADF22A1 are consistent with a luminous, heavily obscured AGN, accompanied by intense, obscured star formation. These results are consistent with the identification of ADF22A1 as a sub-millimeter galaxy hosting an obscured ``proto-quasar.''}
\end{figure*}

ADF22A1 sits at the opposite end of the typical picture of AGN evolution from J221720.24+002019.3. The ADF22A1 proto-quasar system was identified as an AzTEC 1.1 mm source in \citet{tamura2010}. Further study showed that the system consists at least two optical/NIR sources offset from a millimeter and X-ray source, which is not detected in bands up to a few microns. The smoothed FWHM of the SSA22 photometry we use is larger than the projected offset between the galaxies and the millimeter/X-ray source, so we fit the SED of the entire ADF22A1 system. We show the best-fit model and the sampled posterior SFH in \autoref{fig:ADF22A1_sed}. The fit from our models is consistent with the picture of ADF22A1 harboring a heavily obscured/Compton-thick AGN, with $N_H = 5.19^{+3.24}_{-3.13} \times 10^{24}\ {\rm cm^{-2}}$, $\cos i = 0.33^{+0.21}_{-0.23}$, and $\tau_{V,\rm Diff} = 2.9^{+0.1}_{-0.1}$. The SFH suggests an initial burst of star formation in the first $\sim 1$ Gyr after formation, followed by an ongoing, more intense (by a factor of $\sim 10$) burst of star formation in the past $100$ Myr. We measure the average star formation over the past $100$ Myr as $524^{+276}_{-258}\ {\rm M_{\odot}\ yr^{-1}}$, a factor of $\sim 2$ smaller than the estimate from the 1.1 mm flux, but still more than twice the SFR of any of the other X-ray selected AGN in our sample. The mass we recover for the system, $2.00^{+1.16}_{-0.74} \times 10^{11}\ {\rm M_{\odot}}$, is consistent with the high end of the mass estimates from \citet{tamura2010}, $8.2^{+9.0}_{-1.5} \times 10^{10}\ {\rm M_{\odot}}$. The optical attenuation above is also consistent with the value reported in \citet{tamura2010} ($\tau_V = 3.1^{+0.3}_{-0.2}$), and with the level of $V-$band attenuation typically measured in SMGs. We note again that our model does not include reflection, and as such the uncertainties on our estimate of $N_H$ are large, though we find that ADF22A1 is likely Compton-thick, as also found by \citet{tamura2010}. As previously mentioned, Compton-thickness renders our estimates of the AGN parameters (particularly $\dot m$) more uncertain.

The IR emission from the AGN remains highly uncertain due to the lack of mid-IR photometry; JWST observations with MIRI will not only better constrain the AGN IR emission, but allow the resolution of the system into its separate components, including the central X-ray point source, allowing us to separately characterize each component.

%%% -----------------------------------------
%%% Figure: Star forming main sequence
%%% -----------------------------------------
\begin{figure}[h]
\centering
\includegraphics[width=0.5\textwidth]{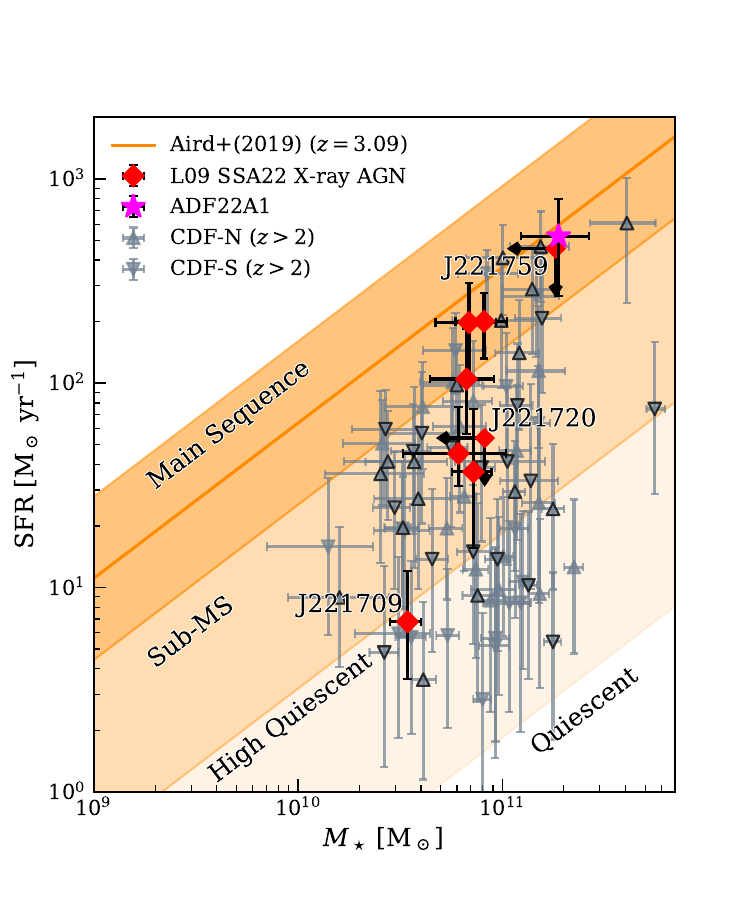}
\caption{\label{fig:main_sequence}We show the SFR averaged over the last 100 Myr plotted against the total stellar mass for the SSA22 X-ray AGN, ADF22A1, and CDF AGN hosts with $z>2$. The orange line shows the redshift-dependent star-forming main sequence from \citet{aird2019} at $z=3.09$, with the shaded regions marking the categories of star formation adopted in that work: main sequence (MS; $\rm -0.4 < \log SFR/SFR_{MS} \leq 0.4$), Sub-MS ($\rm -1.3 < \log SFR/SFR_{MS} \leq -0.4$), high quiescent ($\rm -2.3 < \log SFR/SFR_{MS} \leq -1.3$), and quiescent ($\rm \log SFR/SFR_{MS} \leq -2.3$). We find that 3 of the 6 \citet{lehmer2009a} X-ray AGN for which we constrain the SFH are categorized as sub-MS or below. The two AGN for which we cannot constrain the SFH are labeled and shown as 99th percentile upper limits. The lowest-SFR AGN in our SSA22 sample is also labeled. The CDF galaxies with spectroscopic redshifts are shown with dark-outlined symbols, and the galaxies with photometric redshifts are shown without dark outlines.}
\end{figure}

% -----------------------------------------
% 5. Discussion
% -----------------------------------------
\section{Discussion}
\label{sec:discussion}

In \autoref{sec:results} we presented the results of our SED fits to the SSA22 AGN, finding that they are typically obscured main sequence galaxies. In the following sections, we compare the measured properties of the SSA22 AGN to the field AGN, in order to establish whether there are measurable differences in the properties of the protocluster AGN-host galaxy systems that could drive the observed AGN enhancement. We also attempt to connect the properties of the AGN-host galaxy systems to their local environment, and finally to place our results in the context of the ongoing question of what drives the AGN enhancement in the protocluster.

%%% -----------------------------------------
%%% Table: KS test results
%%% -----------------------------------------
\begin{deluxetable}{l c c c c}
\tablecolumns{5}
\tablecaption{\label{tab:KS_tests} 1D KS test $p-$values for SED-fit derived quantities between SSA22 and GOODS.}
\tablehead{\colhead{} & \multicolumn{2}{c}{$2 \le z \le 4$\tablenotemark{a}} & \multicolumn{2}{c}{$2.6 \le z \le 3.4$}\\ \cmidrule(lr){2-3} \cmidrule(lr){4-5} \colhead{Quantity} & \colhead{No CT\tablenotemark{b}} & \colhead{CT Included\tablenotemark{c}} & \colhead{No CT} & \colhead{CT Included}}
\startdata
$M_\star$          &    \nodata &    0.086 &    \nodata &    0.114 \\
SFR                &    \nodata &    0.286 &    \nodata &    0.232 \\
$M_{\rm SMBH}$     &      0.094 &    0.013 &      0.205 &    0.104 \\
$\rm SFR/SFR_{MS}$ &      0.984 &    0.716 &      0.990 &    0.351 \\
$\rm BHAR/SFR$     &      0.489 &    0.276 &      0.735 &    0.599 \\
$N_H$              &      0.480 &    0.210 &      0.998 &    0.300 \\
% $N_H$            &      0.480 &    0.581 &      0.998 &    0.765 \\
\enddata
\tablecomments{ADF22A1 is excluded from all KS tests; J221720.24+002019.3 and J221759.23+001529.7 are excluded from the KS tests on $M_\star$, $\rm SFR/SFR_{MS}$, and $\rm BHAR/SFR$ as we cannot effectively constrain their SFH.}
\tablenotetext{a}{Redshift range of field galaxies.}
\tablenotetext{b}{$p-$values where galaxies with $N_H \ge 1\times 10^{24}$ are excluded from the KS test.}
\tablenotetext{c}{$p-$values where galaxies with $N_H \ge 1\times 10^{24}$ are included in the KS test.}
\end{deluxetable}

%% -----------------------------------------
%% 5.X Star Formation in the SSA22 AGN hosts
%% -----------------------------------------
\subsection{Star Formation and Black Hole Growth in the Protocluster Active Galactic Nuclei}

\begin{figure}
\centering
\includegraphics[width=0.5\textwidth]{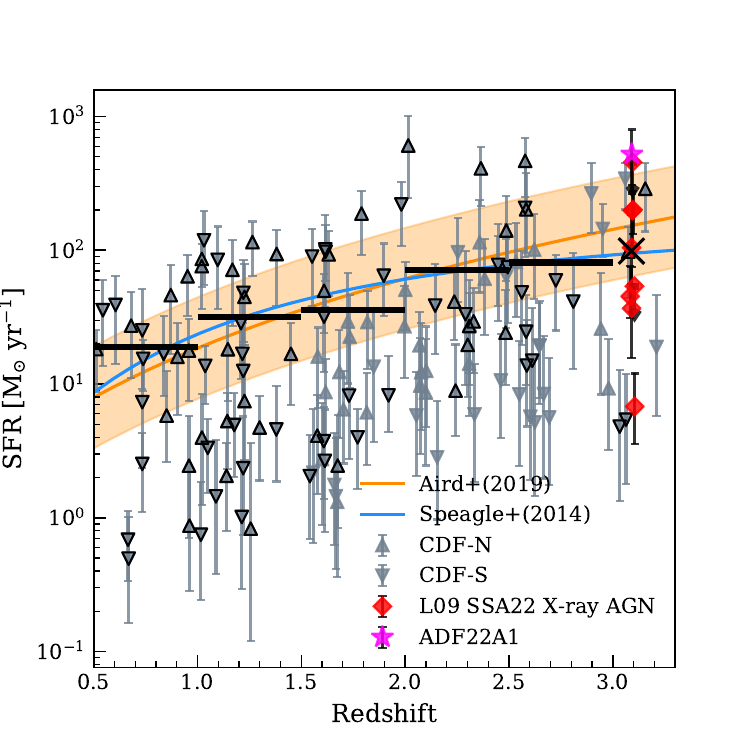}
\caption{\label{fig:sfr_vs_redshift}We show the SFR of all of our galaxies as a function of redshift, up to $z = 3.3$. CDF galaxies with secure spectroscopic redshifts are shown with dark-outlined symbols. Each galaxy's SFR is averaged over the last 100 Myr of the sampled posterior SFH. The mean SFR of the CDF galaxies is shown by the thick black bars over four ranges of redshift, and the mean SFR of the \citetalias{lehmer2009a} X-ray AGN is shown as a black cross. The two AGN for which we cannot constrain the SFH are shown as 99th percentile upper limits. In the background we show the evolution of the star forming main sequence according to \citet{aird2019} and \citet{speagle2014} for a stellar mass of $10^{10.5}\ \rm M_{\odot}$, with the orange shaded region indicating the range of the \citet{aird2019} main sequence for stellar masses $10^{10}-10^{11}\ \rm M_{\odot}$. The average SFR of the SSA22 protocluster AGN is consistent with both the average SFR of the field AGN and the main sequence at $z=3.09$.}
\end{figure}

\citet{alexander2016} previously found, on the basis of IR scaling relations, that the range of SFR of the three IR-detected \citetalias{lehmer2009a} X-ray AGN is consistent both with AGN in field environments and the overall evolution of the star forming main sequence. On average, we also find that the X-ray detected protocluster AGN hosts are consistent with the star forming main sequence at $z=3.09$. In \autoref{fig:sfr_vs_redshift} we show the evolution of the main sequence according to both the \citet{speagle2014} and \citet{aird2019} prescriptions, along with the SFR of all the GOODS AGN in our sample, up to $z=3.3$. We also overlay the SFR of the SSA22 AGN and the sample mean SFR. We find that the average SFR of the \citetalias{lehmer2009a} X-ray AGN in SSA22 is consistent with both the field AGN in the GOODS fields and the main sequence. While our results suggest that $4/8 = 50 \pm 18 \%$ of the \citetalias{lehmer2009a} X-ray AGN are located below the main sequence (as previously observed in the ADF22-QG1 protocluster AGN by \citealp{kubo2021, kubo2022}), we find that sub-main-sequence evolution is just as common in our sample of $L_X$-selected AGN in the CDF: of the $z>2$ ($2.4\leq z \leq 3.6$) AGN, $49/68 = 72 \pm 10\%$ ($15/21 = 71 \pm 18\%$) have $\log\rm SFR/SFR_{MS} \leq -0.4$. This is consistent also with \citet{aird2019}, which finds a slight preference of X-ray selected AGN to lie below their star-forming main-sequence (see e.g., their Figure 8). The fractions of putative quiescent galaxies with $\log\rm SFR/SFR_{MS} \leq -1.3$ is $1/8 = 12.5 \pm 12.5 \%$ for the \citetalias{lehmer2009a} X-ray AGN, and $15/68 = 22 \pm 6 \%$ ($8/21 = 38 \pm 13 \%$) for the $z>2$ ($2.4\leq z \leq 3.4$) CDF AGN. We show Kolmogorov-Smirnov (KS) tests on the posterior distributions of SFR and $\rm SFR/SFR_{MS}$ in \autoref{tab:KS_tests}, suggesting that the SSA22 AGN are consistent with being drawn from the same underlying population distribution as the high-redshift CDF AGN. We find also that the AGN host galaxies in the protocluster are no more massive on average than AGN hosts in the field with similar X-ray luminosities: excluding the ADF22A1 system and the galaxies for which the AGN-only model is preferred, we find that the ratio of the average stellar mass of the protocluster AGN to the average mass of the GOODS AGN is at most 0.945 ($3\sigma$ upper limit) over the redshift range $2 \leq z \leq 4$. The CDF comparison samples naturally probe a much larger volume than the SSA22 AGN sample; as such we might expect to find a few more high-mass galaxies in the field sample. When we restrict redshift to $2.6 \leq z \leq 3.4$ we find that the mass ratio is at most 1.020.

%%% -----------------------------------------
%%% Figure: SFR vs Redshift a la Alexander+(2016) Fig 3
%%% -----------------------------------------

\begin{figure}
\centering
\includegraphics[width=0.5\textwidth]{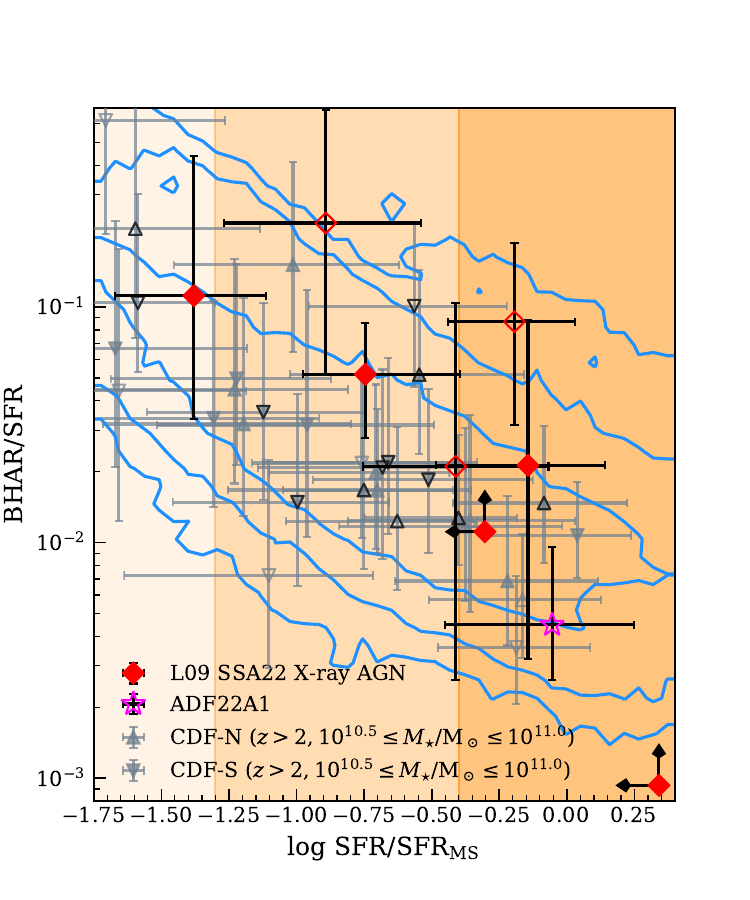}
\caption{\label{fig:BHAR_vs_sfr_ratio} We show the $\rm BHAR/SFR$ ratio as a function of $\rm SFR/SFR_{MS}$ for the SSA22 sample and the field samples. The field samples are shown as points for galaxies with $10^{10.5} \leq \log (M_{\star} / {\rm M_{\odot}}) \leq 11.0$ (the range of stellar mass for SSA22), while the full set of CDF AGN candidates with $z \geq 2$ and $N_H < 10^{24}\ {\rm cm^{-2}}$ are shown as contours, containing $68\%$, $95\%$, and $99\%$ of the sample's posterior probability. The vertical shaded regions are the \citet{aird2019} $\rm SFR/SFR_{MS}$ classifications: from left to right, high quiescent, sub-MS, and MS. CT candidates are shown with unfilled symbols, and field galaxies with spectroscopic redshifts are shown with dark outlined symbols. For the two SSA22 AGN where the AGN-only model is preferred, we show the 99th percentile of $\rm SFR/SFR_{MS}$ as an upper limit and the 1st percentile of $\rm BHAR/SFR$ as a lower limit. The trend in the SSA22 data is well captured by the trend in the field galaxies of the same mass.}
\end{figure}

To further establish the relationship between star formation and the black hole growth in the protocluster, we calculated the dimensionless black hole accretion rate (BHAR) to SFR ratio, $\rm BHAR / SFR$, for each of our samples. This ratio traces the evolutionary state of the black hole relative to the host galaxy, and may shed light on the accretion history of the AGN, as the BHAR is shown to lag behind the SFR following major accretion episodes, as gas is driven down to the small scales near the nucleus where it can be accreted by the SMBH. $\rm BHAR / SFR$ is found to be independent of redshift when controlled for mass \citep[e.g.][]{aird2019}. We calculated the BHAR of our galaxies as ${\rm BHAR} = \dot m \dot M_{\rm Edd}$, with

\begin{equation}
	\dot M_{\rm Edd} = \frac{4 \pi G M m_p}{\sigma_T \eta c},
\end{equation}
and where $m_p$ is the proton mass, $\sigma_T$ is the Thomson scattering cross section, and $\eta$ is the radiative efficiency of the BH. We take $\dot m$ and $M$ from our SED fits. Following \citet{kubota2018}, we assume a radiative efficiency of $\eta = 0.057$. In \autoref{fig:BHAR_vs_sfr_ratio}, we show the $\rm BHAR/SFR$ as a function of $\rm SFR/SFR_{MS}$, where $\rm SFR_{MS}$ is computed at the appropriate redshift and stellar mass for each galaxy using the redshift-dependent main sequence from \citet{aird2019}. We find that the SSA22 AGN follow the same trend in $\rm BHAR/SFR$ as a function of $\rm SFR/SFR_{MS}$ as the field galaxies, indicating that, on the timescales we probe, black hole growth in the protocluster is not proceeding out of pace with galaxy growth at any stage of galaxy evolution. A 1D KS test on the distributions of $\rm BHAR/SFR$ indicates that protocluster and field galaxies have consistent distributions of $\rm BHAR/SFR$ regardless of whether we exclude CT candidates ($p=0.489$) or include them ($p=0.276$) or whether we limit the test to galaxies within the stellar mass range of the SSA22 AGN, $10.5 \leq \log (M_{\star} / {\rm M_{\odot}}) \leq 11.0$ ($p = 0.268$ without CT candidates and $p = 0.161$ with CT candidates). The stochasticity of AGN accretion produces significant scatter in $\rm BHAR/SFR$; as such it is often useful to calculate the sample average, $\langle \rm BHAR/SFR \rangle$. We find $\langle \rm BHAR/SFR \rangle = 0.075_{-0.037}^{+0.111}$ for the SSA22 AGN, when we exclude CT candidates. For the CDF samples, $\langle \rm BHAR/SFR \rangle$ is consistent with SSA22: $0.075_{-0.016}^{+0.039}$ over the whole range of mass and $z \geq 2$, and $0.056_{-0.017}^{+0.031}$ over $10.5 \leq \log (M_{\star} / {\rm M_{\odot}}) \leq 11.0$ and $z \geq 2$, both excluding CT candidates.

%%% -----------------------------------------
%%% Figure: BHAR/SFR space
%%% -----------------------------------------

\begin{figure}
\centering
\includegraphics[width=0.5\textwidth]{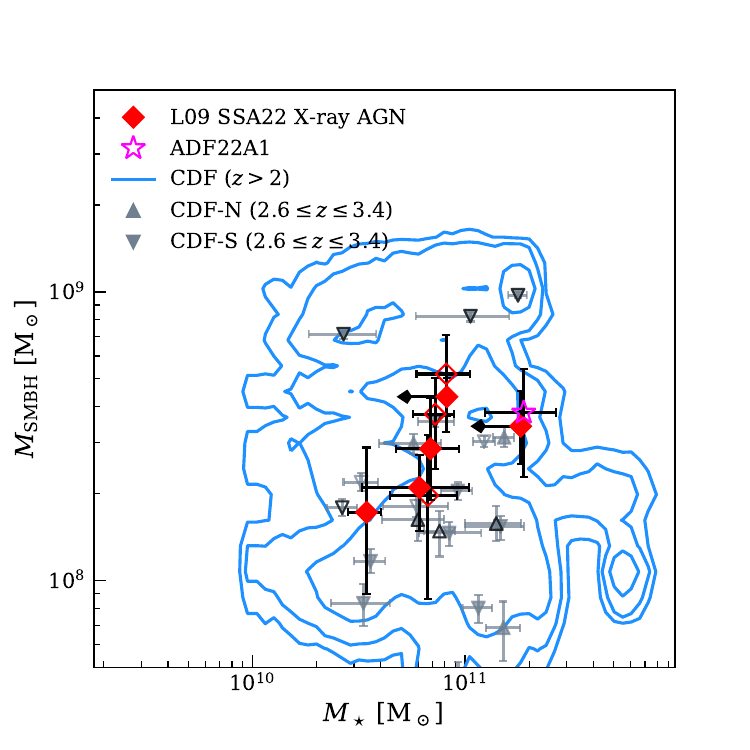}
\caption{\label{fig:bh_mass} We show the SMBH mass of our sample galaxies as a function of stellar mass. As before, unfilled symbols represent CT candidates and CDF symbols with black outlines represent galaxies with spectroscopic redshifts. The $z\geq2$ CDF galaxies with $N_H <10^{24}\ {\rm cm^{-2}}$ are shown as contours containing $68\%$, $95\%$, and $99\%$ of the sample's posterior probability. For the two SSA22 AGN where the AGN-only model is preferred, we show the 99th percentile of the stellar mass as an upper limit. The SSA22 AGN and the CDF AGN with $2.6 \leq z \leq 3.4$ occupy the same region of the $M_{\rm SMBH}$--$M_{\star}$ plane, and the SSA22 AGN occupy a high-probability region of the $z\geq2$ posterior, indicating that black holes have not grown out of pace with their hosts in the protocluster.}
\end{figure}

Another possible explanation for the observed AGN fraction enhancement in the protocluster is that the black holes in protocluster AGN may simply be more massive than their field counterparts. In \autoref{fig:bh_mass} we show the black hole mass of our samples plotted against the stellar mass. Our high-redshift galaxy samples are likely not bulge-dominated (\citet{monson2021} found that none of the SSA22 X-ray AGN are bulge-dominated by $G$--$M_{20}$ diagnostics, and that the AGN that are not point-like sources in F160W typically have S\'ersic indices $n<2$), and as such we lack estimates of the bulge mass. Thus, we are unable to correlate stellar masses strongly with BH masses. We find, however, that the $z>2$ CDF AGN and SSA22 AGN occupy the same general region of the $M_{\rm SMBH}$--$M_{\star}$ plane, and that the $2.6 \leq z \leq 3.4$ CDF AGN track very closely with the SSA22 AGN. This is shown also by the KS test results in \autoref{tab:KS_tests}; when we limit the field sample to a narrower range of redshifts around $z = 3$, we find that the SMBH mass distributions of the SSA22 and GOODS samples are consistent, with $p > 0.104$. We additionally computed the posterior distributions on the sample-averaged black hole mass (excluding CT candidates), finding that the $3\sigma$ upper limit (computed as the 99th percentile of the posterior) on the ratio of the average protocluster SMBH mass (for the 8 \citetalias{lehmer2009a} AGN) to the average field SMBH mass is 1.43 for field galaxies with $2.6 \leq z \leq 3.4$. While empirical SBMH mass to stellar mass relationships for local bulge-dominated galaxies may not be applicable to our sample, the recent FOREVER22 suite of SSA22-like protocluster simulations \citep{yajima2022} allow us a theoretical point of comparison: we find that our estimates of the SMBH masses of our AGN sample are consistent with the SMBH to stellar mass relationship produced by their sample \citep[see Fig. 11 in][]{yajima2022}. Finally, we caution that our estimates of the black hole mass are (by necessity, for high-$z$ AGN) indirect, based on the theoretical \texttt{qsosed} AGN model, and effectively scale with the AGN luminosity. Thus with weak constraints on the AGN luminosity (as in cases where we have few X-ray counts and no mid-IR observations) the estimated black hole masses may be less reliable.

%%% -----------------------------------------
%%% Figure: Black hole mass plot
%%% -----------------------------------------

To examine whether the comparison between the SMBH properties of the SSA22 and CDF AGN is biased by the lack of MIPS $24\ {\rm \micron}$ fluxes (which constrain the NIR AGN torus emission over a wide range of redshift) in the SSA22 SED fits (excluding ADF22A1), we re-fit the field sample without $24\ {\rm \micron}$ measurements. We find that the Eddington ratio parameter $\log \dot m$ is the only parameter to show a systematic difference between the two cases: the fits without MIPS have systematically smaller Eddington ratios than those with MIPS. The median offset in $\log \dot m$ is less than 0.3 dex for $\log \dot m < -1$, increasing to $\approx 1.0$ dex as $\log \dot m$ approaches 0.0. However, the uncertainties on $\log \dot m$ also increase with $\log \dot m$, such that the systematic uncertainty is less than the random uncertainties; we find that the difference between the two measurements is consistent with 0.0 within $1\sigma$ for the majority of systems. While it is thus likely that we underestimate the Eddington ratio (and consequently BHAR) for some of the SSA22 systems, for which we do not use $24\ {\rm \micron}$ measurements, we expect the systematic offset to be smaller than the random uncertainties on our measurements of $\log \dot m$ and the additional uncertainties imposed by the heavy obscuration of the highly accreting SSA22 AGN systems. Our overall results should thus be insensitive to the lack of MIPS $24\ {\rm \micron}$ constraints in our SSA22 SED fits.

%% -----------------------------------------
%% 5.X Environment
%% -----------------------------------------
\subsection{The Possible Role of Local Environment}
\label{sec:environment}

The SSA22 protocluster is well-suited to attempts at linking galaxy properties to the protocluster environment, largely due to the extent to which the cold-gas features of the intergalactic medium (IGM) have been mapped in the protocluster. A number of star-forming galaxies and AGN in the protocluster are associated with Lyman-$\alpha$ blobs (LABs), large scale Ly$\alpha$ emission nebulae \citep{matsuda2004,geach2009}, where the Ly$\alpha$ emission is believed to be powered by heating from the galaxies embedded within \citep{geach2009}. The LABs are in turn associated with the large-scale Ly$\alpha$ emission filaments imaged with MUSE in \citet{umehata2019}. These reservoirs of cold gas in the IGM could play a role in maintaining the levels of star formation and black hole accretion in the protocluster galaxies by providing a steady supply of cold gas for the galaxies within.

Four of the X-ray AGN, J221739.08+001330.7, J221735.84+001559.1, J221759.23+001529.7, and J221732.00+001655.6, are associated with LABs. These AGN do not clearly deviate from the other AGN in the SSA22 sample in terms of any of the parameters or quantities derived from our model. However, two, J221739.08+001330.7 and J221732.00+001655.6, have the highest $N_H$ of the 8 \citetalias{lehmer2009a} X-ray AGN. While \citet{geach2009} also previously noted that the AGN associated with LABs are heavily obscured, whether or not this indicates a greater cold gas density in LAB-associated AGN would require a larger sample and more detailed X-ray modeling, beyond the scope of this work. J221736.54+001622.6 and ADF22A1 (two of the most intrinsically X-ray luminous sources in our sample) are also notably associated with nodes of the IGM filaments. 

While it is believed that the Ly$\alpha$ emission from these nebulae is likely powered by the AGN embedded within them, it remains unclear whether the reservoirs of cold gas in the IGM traced by the filaments and LABs play any role in the fueling of the AGN. However, the LABs and the large-scale filaments are also consistently associated with star formation - the SMGs detected in \citet{umehata2014} also co-locate with the filaments, and each of the LABs associated with an AGN in our sample contains an ALMA Band 7 detection, either directly associated with the AGN or offset \citep{alexander2016}, so it remains likely that the IGM gas reservoirs play some role in supplying the galaxies embedded within them with a supply of cold gas. 

%%% -----------------------------------------
%%% Figure: Host galaxy props and LAE density
%%% -----------------------------------------
\begin{figure}
\centering
\includegraphics[width=0.5\textwidth]{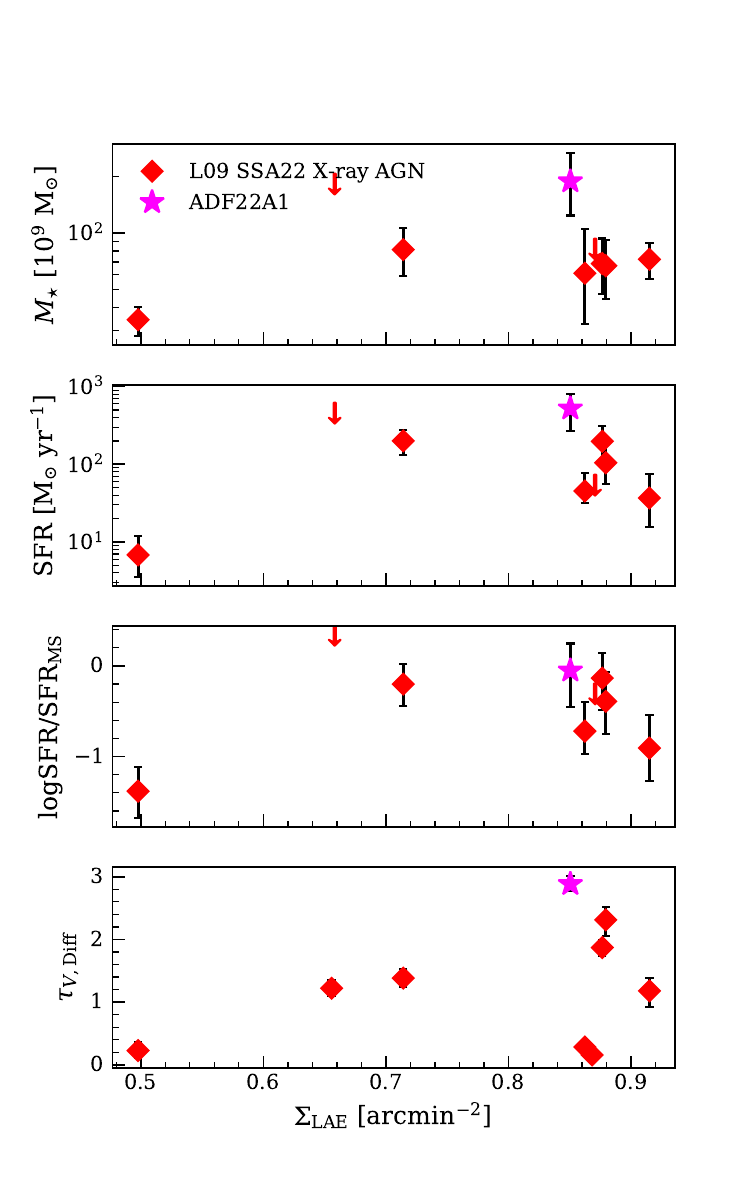}
\caption{\label{fig:host_LAE}We show the SSA22 AGN sample host galaxy properties as a function of the projected surface density of $z\approx3.09$ LAEs from \citet{hayashino2004}. For the two AGN where the AGN-only model is preferred, we show the 99th percentile upper limit for the SFH-derived properties. There are no significant trends observed with the surface density.}
\end{figure}

%%% -----------------------------------------
%%% Figure: AGN props and LAE density
%%% -----------------------------------------
\begin{figure}
\centering
\includegraphics[width=0.5\textwidth]{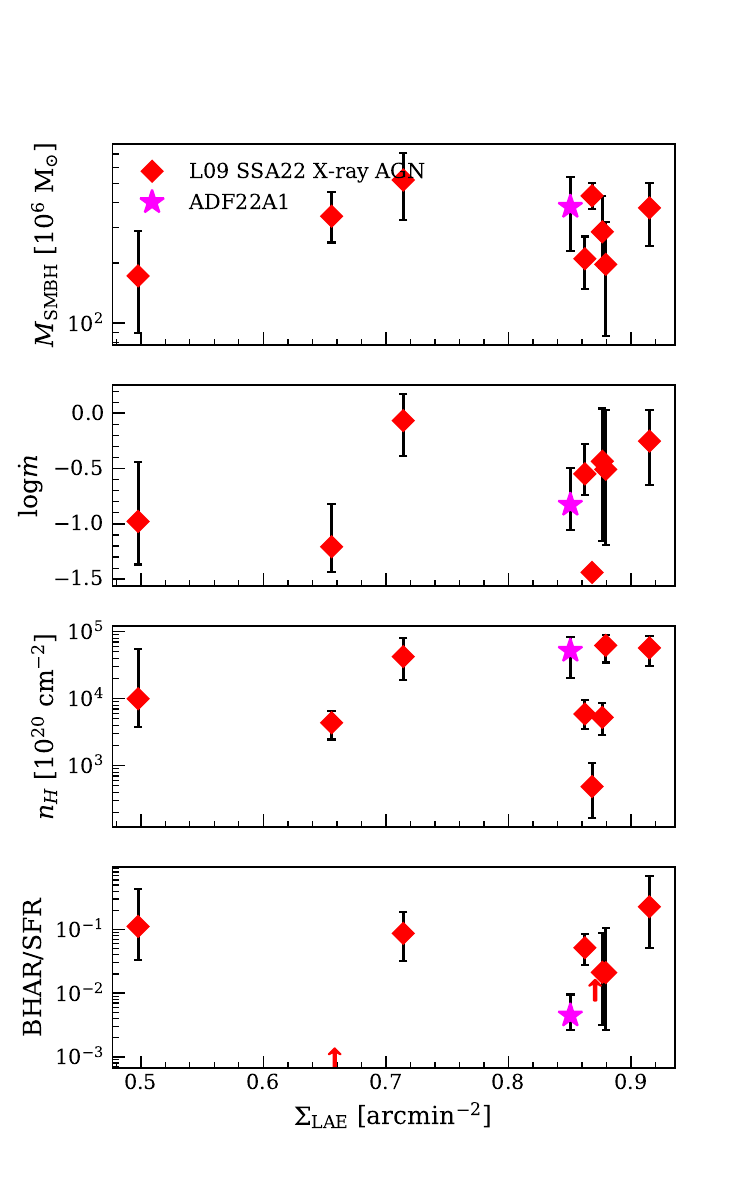}
\caption{\label{fig:AGN_LAE}We show the SSA22 AGN sample SMBH properties as a function of the projected surface density of $z\approx3.09$ LAEs from \citet{hayashino2004}. For the two AGN where the AGN-only model is preferred, we show the 1st percentile lower limit on BHAR/SFR. There are no significant trends observed with the surface density.}
\end{figure}

To attempt to correlate the properties of our AGN sample with their location in the protocluster, we calculated $\Sigma_{\rm LAE}$, the projected surface density of $z\approx3.09$ LAEs from \citet{hayashino2004}, using the same procedure as in \citet{monson2021} with a Gaussian kernel density estimate with a $4'\ (7.5\ {\rm cMpc})$ FWHM\footnote{We use a larger FWHM than the $2'$ used in \citet{monson2021}; we found that a KDE with a $2'$ FWHM resulted in overestimates of the LAE density near those AGN which are themselves LAEs.}. We note, however, that the coherent structure of protocluster LAEs also extends by tens of cMpc along the line of sight \citep{matsuda2005}. The projected surface density of LAEs does not capture the 3D structure of the protocluster, and can only be used as a general gauge of the protocluster density. In \autoref{fig:host_LAE} and \autoref{fig:AGN_LAE}, we show the SED-fitting derived parameters for the SSA22 X-ray AGN and ADF22A1 plotted against the surface density. As previously shown \citep[in, e.g., studies of ADF22:][]{kubo2013,kubo2022,umehata2014,umehata2015}, AGN are preferentially found in the denser regions of the protocluster (ADF22A1 is believed to lie near the density peak of the protocluster), though we find that host galaxy properties and SMBH properties are uncorrelated with $\Sigma_{\rm LAE}$. 

We previously attempted to establish the merger states and recent merger histories of the SSA22 X-ray AGN in \citet{monson2021}. In the Gini-$M_{20}$ (calculated based on F160W images) plane, which is commonly used as a merger diagnostic, we found that none of the X-ray AGN have the clumpy, uneven flux distributions associated with recent mergers, and are instead consistent with concentrations of light near the center, as might be expected from mainly star-forming galaxies hosting AGN \citep[see Figure 6 in][]{monson2021}. We also previously employed blind classifications, with our classifiers voting on whether images contained a merger or an isolated galaxy. Based on the presence of close projected companions, 3 of the X-ray AGN (J221736.54+001622.6, J221739.08+001330.7, J221720.24+002019.3) were classified as possible mergers in this way. However, in these cases the redshifts of the projected companions are unknown and the primary galaxies are not exhibiting the disturbed morphologies associated with an interacting companion galaxy. Our updated estimates of the stellar masses for the SSA22 X-ray AGN place them at $M_{\star} = 10^{10.5}$--$10^{11.3}\ {\rm M_{\odot}}$ (in good agreement with the estimates from \citealp{kubo2015}); the EAGLE simulations \citep{mcalpine2020} suggest that at $z>2$ mergers are most effective at triggering SMBH growth at $M_{\star} < 10^{11.0}\ {\rm M_{\odot}}$, so we again find it unlikely that mergers are currently a dominant growth mechanism for SMBHs in the protocluster.

%% -----------------------------------------
%% 5.X So What's Going On?
%% -----------------------------------------
\subsection{What Drives the Active Galactic Nuclei Fraction Enhancement in the Protocluster?} 

We have found that the protocluster environment does not strongly impact the physical properties of the X-ray selected protocluster AGN or their host galaxies, as measured by X-ray-to-IR SED fitting. However, these same X-ray AGN account for a $\approx6-$fold increase in the AGN fraction over the $z\sim3$ field. Given that our results suggest that protocluster AGN and their hosts coevolve in the same way as field AGN, it seems, then, that AGN-triggering events are more common in the protocluster.

SSA22 protocluster galaxies have been shown to be more massive and star forming, on average, than field galaxies, and \citet{monson2021} found that their SFHs indicated more intense or sustained stellar mass buildup at earlier epochs. As stated above, \citet{monson2021} previously found that mergers are no more common among the protocluster population than in the field. However, the constraints on the merger fractions from \citet{monson2021} are weak, and \citet{hine2016} previously found a marginal merger enhancement based on rest-frame UV visual classifications. We also do not have information about the historic merger fraction in the protocluster, and so it remains possible that past mergers at $z > 3.1$ contributed to the buildup of stellar and BH masses, and to the triggering of the observed AGN. Additionally, the IGM in the protocluster, having been imaged in \lya\ emission, is known to be denser than in the field, especially in the vicinity of AGN and rapidly growing sub-mm galaxies. Smooth accretion of cold gas from IGM reservoirs leading to secular AGN triggering thus also remains a compelling explanation for the growth of galaxies and their SMBHs in the protocluster.

% -----------------------------------------
% 6. Summary and Conclusions
% -----------------------------------------
\section{Summary \& Conclusions}
\label{sec:summary}

We have developed a new method for fitting the X-ray to IR SEDs of AGN-hosting galaxies, based on physically-motivated models wherever possible, and implemented it into our SED-fitting code \ligh. We applied our method to the 8 \citetalias{lehmer2009a} X-ray detected protocluster AGN and the candidate protoquasar ADF22A1 in SSA22, and to a comparison sample of 151 X-ray selected candidate AGN at $0.45 \leq z \leq 4.52$ in the \textit{Chandra} Deep Fields.

The normal (i.e., non-AGN) star-forming population of the protocluster has previously been shown to host an enhancement of SFR and stellar mass over star-forming galaxies in the field at the same redshift \citep[e.g.][]{kubo2015, kato2016, monson2021}. However, we find here that the stellar mass and SFR distribution of the X-ray selected protocluster AGN is consistent with the stellar mass and SFR of X-ray selected field AGN, suggesting that AGN hosts in the mass range we probe grow by the same mechanisms in protocluster and field environments. We also find comparable black hole masses and growth rates in the protocluster and field, suggesting normal black hole growth, and thus normal black hole-galaxy coevolution, in the the X-ray selected protocluster AGN. We interpret these results together with the observed AGN fraction enhancement as suggesting that AGN triggering events are more likely in the protocluster. Given the (albeit weakly constrained) lack of merger enhancement found among protocluster LBGs in \citet{monson2021}, this could indicate secular AGN triggering is more common in the protocluster than the field, possibly due to environmental effects related to the dense IGM in the protocluster.

\begin{itemize}
\item{We find that SMBH growth in the SSA22 protocluster is largely obscured: 6/8 X-ray selected AGN are heavily obscured, with $N_H \geq 5 \times 10^{23}\ {\rm cm^{-2}}$, and 3/8 are Compton-thick candidates with $N_H \geq 1 \times 10^{24}\ {\rm cm^{-2}}$. The majority of the protocluster AGN, 5/8, also have optical SED fits consistent with optically obscured AGN growth, with $\cos i \lesssim 0.65$.}
\item{At least 3/8 of the SSA22 protocluster AGN have SFHs consistent with sub-MS growth, with $\log {\rm SFR/SFR_{\rm MS}} \leq -0.4$. J221720.24+002019.3, which is better-fit by an AGN-only model, is also likely located below the main sequence, with an upper-limit $\log {\rm SFR/SFR_{\rm MS}} = -0.31$. One of the sub-MS AGN, J221709.60+001800.1 is identified as a probable quiescent galaxy, with $\log {\rm SFR/SFR_{\rm MS}} = -1.38^{+0.26}_{-0.29}$.}
\item{The protocluster AGN hosts are no more massive than the field AGN hosts in our sample: in a KS test comparing the stellar mass distributions of the SSA22 and CDF samples, we find a $p-$value of at least 0.086, suggesting that the distributions are consistent. Likewise we find a KS $p-$value of at least 0.232 for SFR.}
\item{We find that the protocluster AGN hosts are in similar evolutionary states (relative to the main sequence) when compared to the $L_X$-selected field galaxies: the KS test $p-$value on $\rm SFR/SFR_{MS}$ is at least 0.351. Additionally, when controlled for mass, we find that the sample-averaged BHAR/SFR ratio, measuring the rate of SMBH growth relative to host galaxy growth, is also consistent between protocluster and field samples: $\langle {\rm BHAR / SFR} \rangle = 0.075_{-0.037}^{+0.111}$ for the SSA22 protocluster AGN and $0.056_{-0.017}^{+0.031}$ for the CDF AGN candidates with $10.5 \leq \log (M_{\star} / {\rm M_{\odot}}) \leq 11.0$ and $z \geq 2$.}
\item{Our estimates of the black hole masses of our samples are also consistent ($p > 0.104$) when we limit the field sample to $2.6 \leq z \leq 3.4$. We constrain the sample-averaged black hole mass of the protocluster AGN to at most 1.43 times that of the the field AGN sample.}
\item{For the ADF22A1 protoquasar system, we find a ${\rm SFR} = 524^{+276}_{-258}\ {\rm M_{\odot}}$ and a stellar mass $2.00_{-0.74}^{+1.16} \times 10^{11}\ {\rm M_{\odot}}$. We estimate the black hole mass at $3.98_{-1.47}^{+2.33} \times 10^8\ {\rm M_{\odot}}$, though this estimate is rendered more uncertain by the Compton-thickness of the system; we find $N_H = 5.62_{-2.88}^{+2.89} \times 10^{24}\ {\rm cm^{-2}}$. Our new constraints on the properties of this system from panchromatic SED fitting are consistent with the established picture of ADF22A1 as one of the most obscured and highly star-forming systems in the protocluster.}
\end{itemize}

% ------------------------------------------
% Acknowledgments
% ------------------------------------------
\begin{acknowledgments}
We gratefully acknowledge support from the NASA Astrophysics Data Analysis Program (ADAP) grant 80NSSC20K0444 (EBM, KD, RTE, BDL). This work has made use of the Arkansas High Performance Computing Center, which is funded through multiple National Science Foundation grants and the Arkansas Economic Development Commission. CMH acknowledges funding from an United Kingdom Research and Innovation grant (code: MR/V022830/1). HU acknowledges support from JSPS KAKENHI Grant Number 20H01953.
\end{acknowledgments}

% ------------------------------------------
% Software
% ------------------------------------------
\software{\textsc{ACISExtract}, \citet{broos2010,broos2012};
\textsc{Astropy}, \citet{astropy2013,astropy2018};
\textsc{CIAO}, \citet{fruscione2006};
\textsc{Lightning}, \citet{eufrasio2017,doore2021,doore2023};
\textsc{P\'egase}, \citet{fioc1997,fioc1999}}

% ------------------------------------------
% Appendices
% ------------------------------------------
\clearpage
\appendix

%% -----------------------------------------
%% A1. Alternative SED fit for J221716
%% -----------------------------------------
\section{Alternative SED Fit for J221716.16+001745.8}
\label{app:AGN7_alternative}

%%% -----------------------------------------
%%% Table: SED props from alt. fit for J221716
%%% -----------------------------------------
\begin{deluxetable*}{l r r c r r r r r}
\tablecolumns{9}
\tablecaption{\label{tab:AGN7_properties}SED-fit derived properties for J221716.16+001745.8. The column meanings are the same as \autoref{tab:SSA22_properties}.}
\tablehead{\colhead{ID} & \colhead{$\log M_{\star}$} & \colhead{SFR} & \colhead{${\rm log SFR/SFR_{MS}}$} & \colhead{$\tau_V$} & \colhead{$\cos i$} & \colhead{$\log M_{\rm SMBH}$} & \colhead{$\log \dot m$} & \colhead{$N_H$} \\ \colhead{} & \colhead{$(\rm M_\odot)$} & \colhead{($\rm M_\odot\ yr^{-1}$)} & \colhead{(dex)} & \colhead{} & \colhead{} & \colhead{$(\rm M_\odot)$} & \colhead{} & \colhead{($10^{22}\ \rm cm^{-2}$)}}
\startdata
{\bf J221716.16+001745.8}&    $\bf 10.9^{+0.1}_{-0.2}$&    $\bf 184.8^{+71.1}_{-71.0}$&    $\bf -0.20^{+0.26}_{-0.29}$&    $\bf 1.4^{+0.2}_{-0.2}$&    $\bf 0.30^{+0.15}_{-0.18}$&    $\bf 8.8^{+0.1}_{-0.2}$&    $\bf 0.0^{+0.2}_{-0.3}$&    $\bf 460.35^{+349.95}_{-239.14}$\\
\enddata
\end{deluxetable*}

%%% -----------------------------------------
%%% Figure: SED plot for alt. fit to J221716
%%% -----------------------------------------
\begin{figure*}
\centering
\includegraphics[width=\textwidth]{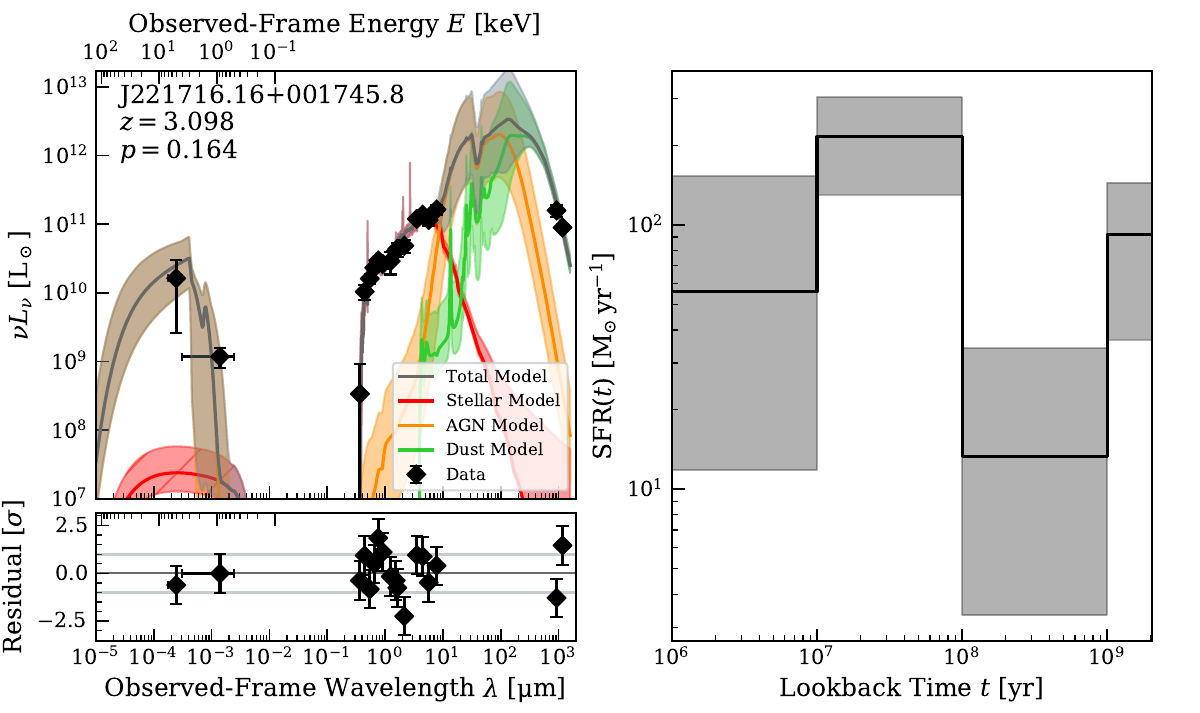}
\caption{\label{fig:AGN7_alternative}We show the alternative SED fit for J221716.16+001745.8. The meanings of the symbols and shaded regions are the same as the previous SED plots. We recover the same SFH shape and SFR when we fit this galaxy with the offset ALMA Band 7 detection, rather than the ALMA Band 7 upper limit at the position of the Chandra detection.}
\end{figure*}

We show the alternative SED fit for J221716.16+001745.8 in \autoref{fig:AGN7_alternative} and the derived properties in \autoref{tab:AGN7_properties}, where we have fit using the ALMA Band 7 offset detection rather than the upper limit from the location of the Chandra detection. Our results for this galaxy are not sensitive to this change, with the exception that we recover a much better fit than when we use the upper limit. In the fit discussed in \autoref{sec:sed:fitting} and \autoref{sec:results}, the uncertainty on the IR contribution of the AGN model (compare the orange bands showing the range of the best $68\%$ of models in \autoref{fig:SSA22_fits_d} and \autoref{fig:AGN7_alternative}) is higher due to the larger uncertainty on the IR emission overall. This results in the alternative fit here having a slightly different $\cos i$ and $\dot m$, though both are consistent with the original fit and the uncertainty in $\dot m$ due to the heavy X-ray obscuration is larger than the shift in $\dot m$ introduced by changing the IR data we use for the fit. The shape of the SFH is only slightly different, in the $10^7-10^8$ yr bin, resulting in a slightly smaller SFR, though well within the uncertainties. 

%% -----------------------------------------
%% A2. AGN-only model fits
%% -----------------------------------------
\section{Active Galactic Nucleus-only Model Fits for SSA22 X-ray AGN}
\label{app:SSA22_pure_AGN}

\begin{figure}
\centering
\includegraphics[width=0.5\textwidth]{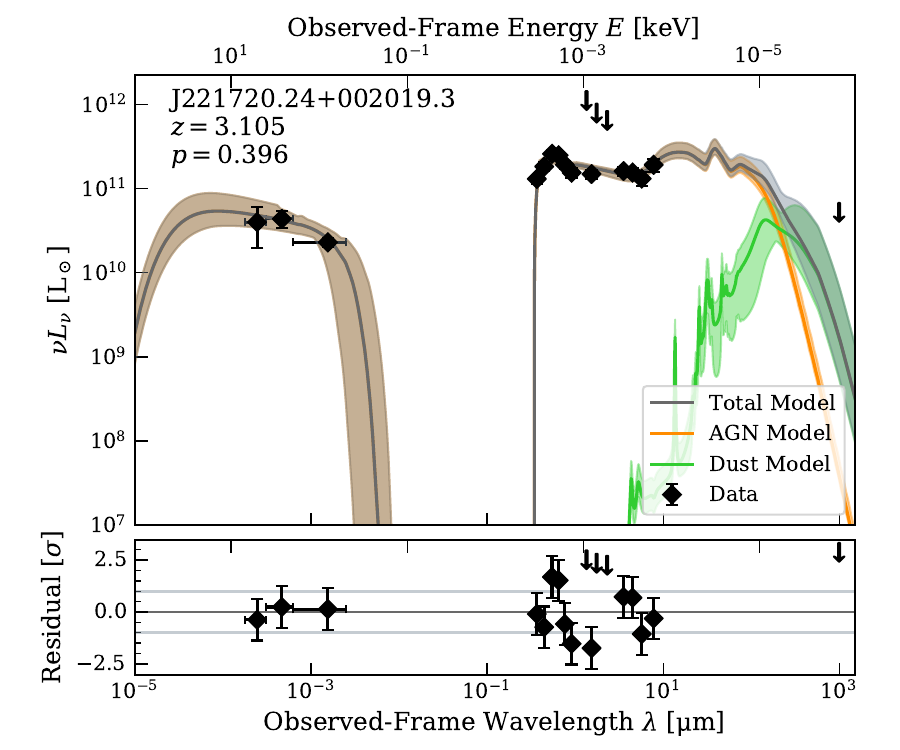}
\caption{\label{fig:AGN4_pureAGN}We show the best-fit AGN-only model for J221720.24+002019.3. All symbols and shaded regions have the same meaning as in the SED plots in \autoref{fig:SSA22_fits_a}.}
\end{figure}

\begin{figure}
\centering
\includegraphics[width=0.5\textwidth]{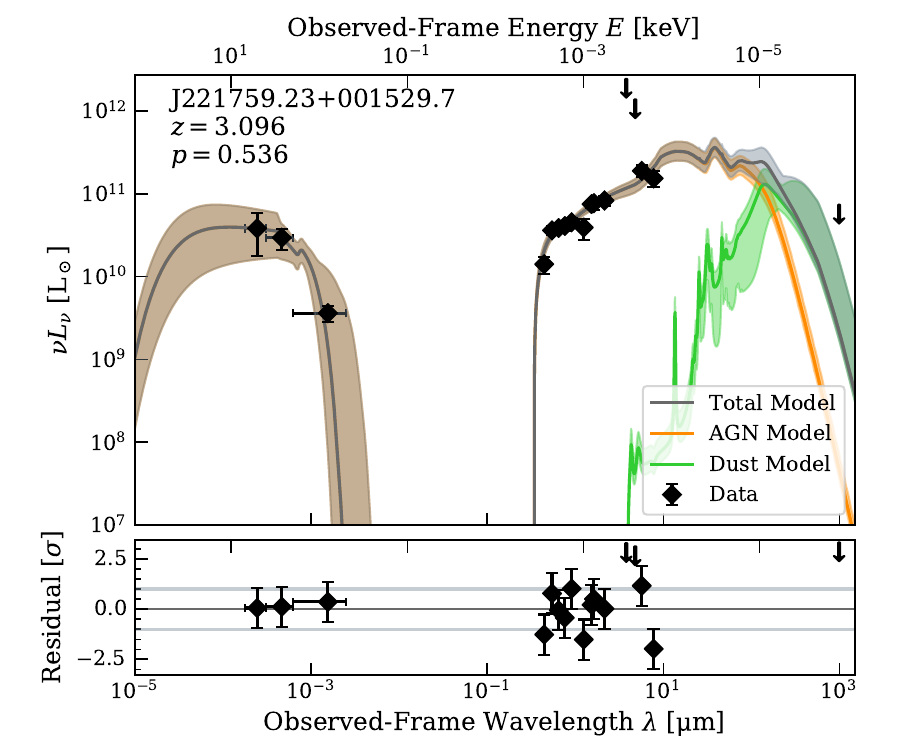}
\caption{\label{fig:AGN6_pureAGN}We show the best-fit AGN-only model for J221759.23+001529.7. All symbols and shaded regions have the same meaning as in the SED plots in \autoref{fig:SSA22_fits_a}.}
\end{figure}

To construct the AGN-only model fits, we set all SFH coefficients in the model to identically 0. The priors, assumptions, and free parameters for the fit are otherwise unchanged from \autoref{tab:SEDParams}. We perform the same PPC goodness-of-fit analysis on these fits as described in \autoref{sec:sed:fitting}. For all but two AGN, J221720.24+002019.3, and J221759.23+001529.7, the AGN-only model is ruled out with $p < 0.001$. The best-fit AGN-only models for these two AGN are shown in \autoref{fig:AGN4_pureAGN} and \autoref{fig:AGN6_pureAGN}.

%%% -----------------------------------------
%%% Table: Model comparison test results
%%% -----------------------------------------
\begin{deluxetable}{l c c c}
\tablecolumns{4}
\tablecaption{\label{tab:pureAGN_stats}AIC values and $F-$test $p-$values comparing models with stellar population emission and AGN-only models.}
\tablehead{\colhead{} & \multicolumn{2}{c}{AIC} & \colhead{} \\ \cmidrule{2-3} \colhead{ID} & \colhead{w/ stellar pop.} & \colhead{w/o stellar pop.} & \colhead{$F-$test $p-$value}}
\startdata
J221720.24+002019.3   & 25.34    & 19.56    & 0.5432 \\
J221759.23+001529.7   & 24.35    & 17.83    & 0.6777 \\
\enddata
\end{deluxetable}

To compare these fits to our original model, we use the Akaike information criterion (AIC) and $F-$test. AIC values and $F-$test $p-$values for the galaxies where the AGN-only model is not ruled out are shown in \autoref{tab:pureAGN_stats}. As a heuristic, the model with lower AIC should be preferred. The $F-$test $p-$value represents the probability of the null hypothesis that the more complex model (in our case, the model including stellar population emission) does \textit{not} provide a significantly improved fit over the simpler model (AGN emission only). We find that the AGN-only model with no stellar population is preferred in both cases.

The recovered $M_{\rm SMBH}$, $\log \dot m$, and $N_H$, are consistent with the original fits. However, to produce the required amount of optical emission, the AGN-only models necessarily have larger $\cos i$, indicating a less-obscured line-of-sight to the accretion disk, than the models including stellar population emission.  While J221720.24+002019.3 is identified in the literature as a spectroscopically confirmed Type 1 AGN, this is not the case for J221759.23+001529.7. The morphology of J221759.23+001529.7 is also extended in F160W, indicating that it is unlikely that AGN continuum emission is dominant across the entire optical-NIR spectrum. Rather, the statistical preference for the AGN-only model may be due to limited NIR constraints.

%%% -----------------------------------------
%%% Figure: AGN-only fit for J221720
%%% -----------------------------------------

%%% -----------------------------------------
%%% Figure: AGN-only fit for J221759
%%% -----------------------------------------

Notably, the AGN for which the AGN-only model is preferred are IR undetected. Our model includes ISM dust heating by the AGN (visible as the green curve in \autoref{fig:AGN4_pureAGN} and \autoref{fig:AGN6_pureAGN}), but heating associated with the AGN produces comparatively little sub-mm emission. For the galaxies in our SSA22 sample with sub-mm detections, the sub-mm constraint gives us an effective handle on the level of star formation, ruling out solutions with zero star formation. J221720.24+002019.3 and J221759.23+001529.7 also have limited NIR constraints. For J221720.24+002019.3 we treat the $JHK$ measurements as upper limits due to possible contamination, such that the stellar emission across the 4000 \AA\ break is poorly constrained. For J221759.23+001529.7 we treat the \spitzer\ IRAC $3.6$ \micron\ and $4.5$ \micron\ measurements as upper limits, due to contamination by a foreground star, preventing us from clearly separating the tail of the stellar population emission from the AGN torus emission.

% ------------------------------------------
% References
% ------------------------------------------
%\clearpage
\bibliographystyle{aasjournal}
\bibliography{SSA22_AGN}

\end{document}